\documentclass[11pt]{article}

\usepackage[a4paper,margin=1in]{geometry}
\usepackage{standalone}
\usepackage{graphicx}
\usepackage{authblk}
\usepackage{caption}
\usepackage{subcaption}
\usepackage{booktabs}
\usepackage{multirow}
\usepackage{multicol}
\usepackage{xcolor}
\usepackage{threeparttable}
\usepackage{amsmath}
\usepackage{amsfonts}
\usepackage{natbib}
\usepackage{array}
\usepackage{hyperref}
\usepackage{enumitem} % Required for customizing list labels
\usepackage{url} 
\newcommand{\Dnew}{\mathcal{D}_0}   % prospective/current trial data
\newcommand{\Dhist}{\mathcal{H}}    % historical dataset
\DeclareMathOperator*{\argmax}{arg\,max}

\title{Selection Bias in Hybrid Randomized Controlled Trials using External Controls: A Simulation Study}

\author{Han Chang Chiam}
\author{Franz K\"onig}
\author{Martin Posch}

\affil{Center for Medical Data Science, Medical University of Vienna, Vienna, Austria}

\date{October 2025}

\begin{document}

\maketitle
\noindent\textbf{Corresponding author:} Prof.\ Martin Posch, Center for Medical Data Science, Medical University of Vienna, Spitalgasse 23, 1090 Vienna, Austria. Email: martin.posch@meduniwien.ac.at

\abstract{Hybrid randomized controlled trials (hybrid RCTs) integrate external control data, such as historical or concurrent data, with data from randomized trials. While numerous frequentist and Bayesian methods, such as the test-then-pool and  Meta-Analytic-Predictive prior, have been developed to account for potential disagreement between the external control and randomized data, they cannot ensure strict type I error rate control. However, these methods can reduce biases stemming from systematic differences between external controls and trial data. A critical yet underexplored issue in hybrid RCTs is the prespecification of external data to be used in analysis.\\

The validity of statistical conclusions in hybrid RCTs depends on the assumption that external control selection is independent of historical trials outcomes. In practice, historical data may be accessible during the planning stage, potentially influencing important decisions, such as which historical datasets to include or the sample size of the prospective part of the hybrid trial, thus introducing bias. Such data-driven design choices can be an additional source of bias, which can occur even when historical and prospective controls are exchangeable.\\

Through a simulation study, we quantify the biases introduced by outcome-dependent selection of historical controls in hybrid RCTs using both Bayesian and frequentist approaches, and discuss potential strategies to mitigate this bias. Our scenarios consider variability and time trends in the historical studies, distributional shifts between historical and prospective control groups, sample sizes and allocation ratios, as well as the number of studies included. The impact of different rules for selecting external controls is demonstrated using a clinical trial example.}


\footnotetext{\textbf{Abbreviations:} ELIR, expected local-information-ratio; EMA, European Medicines Agency; ESS, effective sample size; FDA, United States Food and Drug Administration; hybrid RCTs, Hybrid randomized controlled trials; ICH, International Council for Harmonisation of Technical Requirements for Pharmaceuticals for Human Use; MAP, Meta-Analytic-Predictive; MHRA, United Kingdom Medicines and Healthcare products Regulatory Agency; ODS, outcome-dependent selection; RCT, Randomized controlled trial; RD, risk difference; RMSE, root mean square error; TTP, test-then-pool.}

\section{Introduction}
Randomized controlled trials (RCTs) are the gold standard for establishing the safety and efficacy of new therapies. In rare or life-threatening diseases, conventional RCT may be impractical or even unethical because the required sample sizes are large, expensive, time-consuming, and patients face a substantial placebo burden. Recent open-science initiatives and data-sharing portals \cite{noauthor_clinicalstudydatarequestcom_nodate, noauthor_yoda_nodate, pencina_supporting_2016} now give investigators access to patient data from completed trials, disease registries, and electronic health records. These resources motivate the integration of external information in the decision making of RCT to increase statistical efficiency, reduce sample size, and accelerate the conduct of RCT. Hybrid randomized controlled trials (hybrid RCTs) achieve this by supplementing the concurrently randomized control arm with external controls. Compared to single-arm trials and thresholding approaches \cite{eichler2016threshold}, they preserve certain benefits of randomization and increase the chance that participants receive the novel treatment with unequal randomization \cite{pocock_combination_1976, hueber_secukinumab_2012}. If the observed randomized and external controls differ substantially, the analysis can revert to using only the internal controls, thereby mitigating biases of treatment comparisons \cite{yuan_design_2019, van_eijk_hybrid_2023, guo_adaptive_2024}.

Because external controls can introduce bias, the International Council for Harmonization of Technical Requirements for Pharmaceuticals for Human Use (ICH) E10 guideline states that external controls are acceptable “only in limited circumstances”, typically when a large, clearly attributable treatment effect is expected in a disease with a well-characterized natural history \cite{noauthor_international_2000}. The United States Food and Drug Administration (FDA) Rare-Diseases guidance also mentioned the same principles for serious conditions with predictable outcomes \cite{noauthor_fdacdercber_2023}.

The use of external data in clinical trials has received increasing attention in the past decades, and regulatory agencies have released several guidance documents in this context (EMA 2018 \cite{noauthor_european_nodate}; FDA 2023a, b \cite{noauthor_fdacdercberoce_2023, noauthor_fdacdercberoce_2023-1}). Two very recent drafts reinforce this trajectory: the United Kingdom Medicines and Healthcare products Regulatory Agency (MHRA) 2025 draft guideline on real-world-data external-control arms demands that sponsors prespecify and lock the external comparator set, and the Section 5.3 of ICH E20 draft on adaptive designs similarly emphasizes the prospective justification of any external data incorporated into an adaptive or hybrid design \cite{noauthor_mhra_2025, noauthor_international_2025}. Related issues were also discussed in the European Medicines Agency (EMA) workshop on the use of Bayesian statistics in clinical development \cite{noauthor_european_workshop}, including metrics to assess Bayesian trials and use cases of historical borrowing with Bayesian methods. A planned reflection paper from the EMA on the use of Bayesian statistics in clinical development will outline the required information and justification for employing Bayesian methods, including prespecification and reporting of data sources. Subsequently, a draft concept paper on the development of a reflection paper regarding the use of external controls for evidence generation in regulatory decision-making was published for public consultation \cite{noauthor_european_nodate}. However, the use of external controls specifically for augmentation is not yet covered, and corresponding regulatory documents remain under development. A recent scoping review by Bofill Roig et al. \cite{bofillroig_use_2023} identified 37 EMA, FDA, and ICH guidance documents that mentioned non-concurrent or external controls; the most common circumstances for the use of external control in identified guideline documents were "rare disease" or "indication specific" concerns followed by "large treatment effect" and "ethical concerns".

The use of external controls relies on the exchangeability assumption whereby the distribution of potential outcomes in the external control group is similar to that in the randomized control group, conditional on relevant baseline covariates. A well-acknowledged challenge is the risk of bias when external and internal data are not exchangeable, a situation referred to as a prior-data conflict. Dynamic borrowing techniques have therefore been developed to down-weight or discount external information based on its agreement with the trial data. Examples include adaptive power priors, meta-analytic approaches and model averaging approaches in the Bayesian framework \cite{gravestock_adaptive_2017, neuenschwander_summarizing_2010, kaizer_bayesian_2018}, as well as test-then-pool (TTP), adaptive lasso and recently developed conformal selective-borrowing methods \cite{viele_use_2014, li_frequentist_2023, zhu_enhancing_2024} in the frequentist setting. These methods are more robust than approaches in which the contribution of external data is fixed a priori such as power prior \cite{chen_power_2000}.

Most of these techniques operate at the analysis stage and assume that the external data is fixed in advance. The integrity of statistical conclusions in hybrid RCTs depends on the assumption that external control selection is independent of trial outcomes of the historical control data. In practice, historical data may be available during the planning stage and may influence which datasets are included or how the prospective part of the trial is sized, subsequently introducing bias. While the appropriateness of external data can be assessed using Pocock’s criteria and has been discussed in several studies \cite{pocock_combination_1976, hatswell_summarising_2020, ghadessi_roadmap_2020, lim_reducing_2020, nuno_use_2025}, the process of selecting external controls is complex and depends on a range of factors. When multiple historical trials are available, even if a rational for the choice of historical controls is provided, one may not be able to rule out that the selection of historical controls is also influenced by outcome data of these trials. In particular, if trial results are in the public domain when planning a hybrid trial, the hybrid trial protocol can be written based on this information. Current methodological approaches focus on adjustments to ensure comparability between external data and prospective randomized data. Less attention is given to aspects as the information that is available at the time controls are selected or on ways to measure the potential bias that arises if the choice of historical controls was informed by the outcome of the historical trials. A particular concern is the bias introduced when controls are selected because their outcomes are more negative than expected. Once the selected pool of studies is biased, even robust dynamic-borrowing may then pass at least part of that bias to the final inference, leading to biased hypothesis tests and treatment effect estimates. This outcome-dependent selection (ODS) bias has been identified by both regulators and independent reviewers as a threat to validity; ICH E10 and FDA’s 2023 draft guidance on externally controlled trials both stress that external-control sources must be finalized before any comparative analysis, and Burger et al. labeled such post-hoc choices as “retrospective selection bias”\cite{noauthor_international_2000,noauthor_fdacdercberoce_2023, burger_use_2021}.

To further illustrate, suppose three historical trials are available with observed control response rates of 0.15, 0.20, and 0.25, and a hybrid RCT is planned. Because these outcomes are already known at the design stage, one might be tempted to exclude a trial based on its outcome (i.e., outcome-dependent selection - ODS) to align the historical pool with an anticipated prospective control rate near 0.20. For example, dropping the highest-rate trial (0.25) reduces apparent heterogeneity but also lowers the pooled historical mean to $(0.15+0.20)/2=0.175$, which can tilt borrowing toward larger estimated treatment effects. In practice there are often seemingly reasonable grounds to exclude a study, such as age of data, differences in endpoint definition, or population shifts, but once these decisions are linked to observed outcomes they constitute ODS and might introduce bias.

To assess the risk of ODS, we carry out a comprehensive simulation study that quantifies the bias introduced when historical trials are selected with knowledge of their outcomes and investigates how that bias varies with study-level heterogeneity, the size and number of available historical studies. We compare Bayesian and frequentist dynamic borrowing analyses and evaluate practical strategies for mitigating the resulting distortion. Importantly, the selection algorithms we study are not recommended to be used in practice. They are intended to mimic real-world scenarios in which sponsors might (deliberately or inadvertently) trim the external-control pool for a more favorable conditional power before submitting the statistical analysis plan. We show that both Bayesian and frequentist methods are prone to such selection bias, and the various selection rules investigated in the simulation studies illustrate how severe the bias may potentially become.

Two simulation frameworks are commonly used to evaluate historical borrowing methods. In a \emph{conditional} framework, the historical dataset is treated as fixed and only the prospective RCT is evaluated, so operating characteristics (e.g., type I error rate, power) are interpreted relative to the observed external evidence. In an \emph{unconditional} framework, both historical and prospective data are generated in each replicate, and results average over variability in both sources, reflecting long-run performance across possible historical evidence pools. The simulation study in this paper adopts the unconditional framework. By contrast, the case study is conditional: a fixed historical pool is used to (i) evaluate design operating characteristics via exact calculations for binary endpoints and (ii) perform posterior analyses given the observed trial data.

The remainder of the paper is organized as follows. Section 2 describes the analysis framework. We will introduce both Bayesian and frequentist methods to allow statistical inference while utilizing (some) historical control data. Section 3 describes the simulation setup and reports the operating characteristics of different selection rules from the simulation. A set of selection rules of the historical controls are explored to investigate potential biases in an extensive setup for clinical trial simulations. Section 4 illustrates the impact of the selection rules with a case study. Section 5 discusses implications for protocol development and offers recommendations on prespecifying external-data selection to minimize outcome-dependent selection bias.

\section{Methods}\label{Methods} 

\begin{figure}[!htbp]
\centerline{\includegraphics[width=\textwidth]{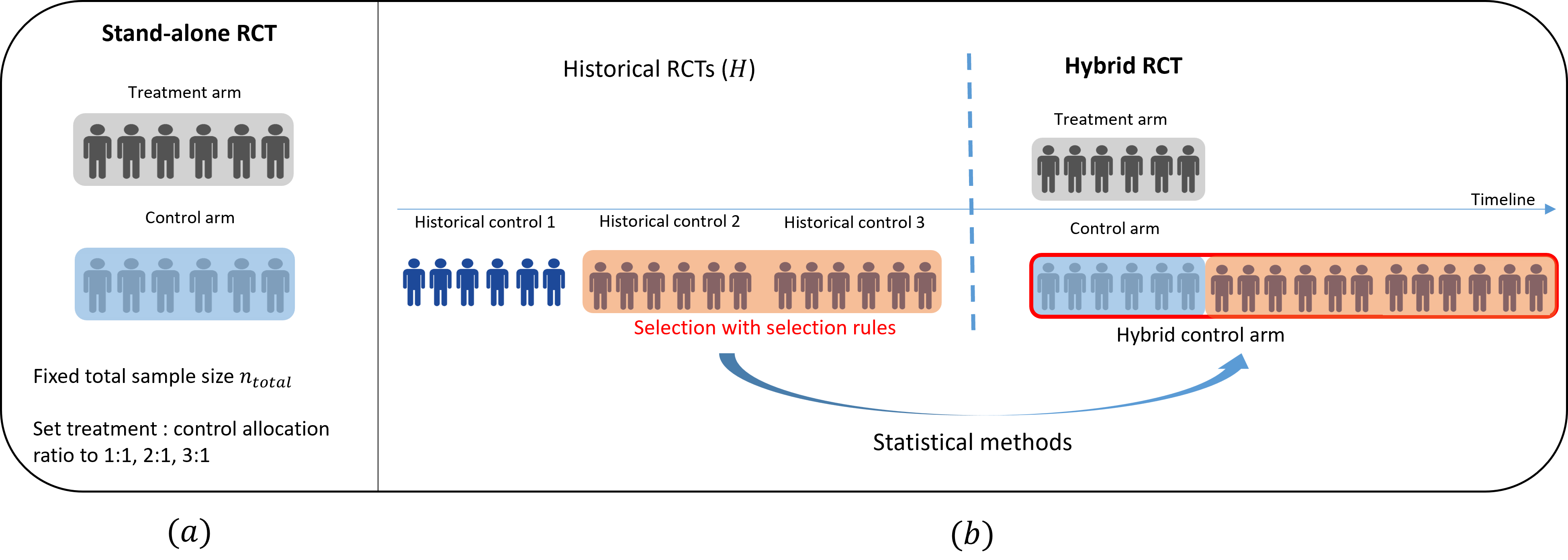}}
\caption{(a) Stand-alone RCT with prospective treatment arm in gray, prospective control arm in blue (b) Illustration of the selection of historical trials and the structure of a hybrid RCT. The x-axis shows a timeline, the vertical blue dashed line differentiates the historical RCTs data and the prospective hybrid RCT. Selected historical trials are shown in orange, the prospective treatment arm in gray, and the prospective control arm in blue. Selected historical controls are combined with the prospective control to form a hybrid control arm (red circle).}
  \label{fig:hybrid_RCT}
\end{figure}

\begin{figure}[!htbp]
\centerline{\includegraphics[width=0.8\textwidth]{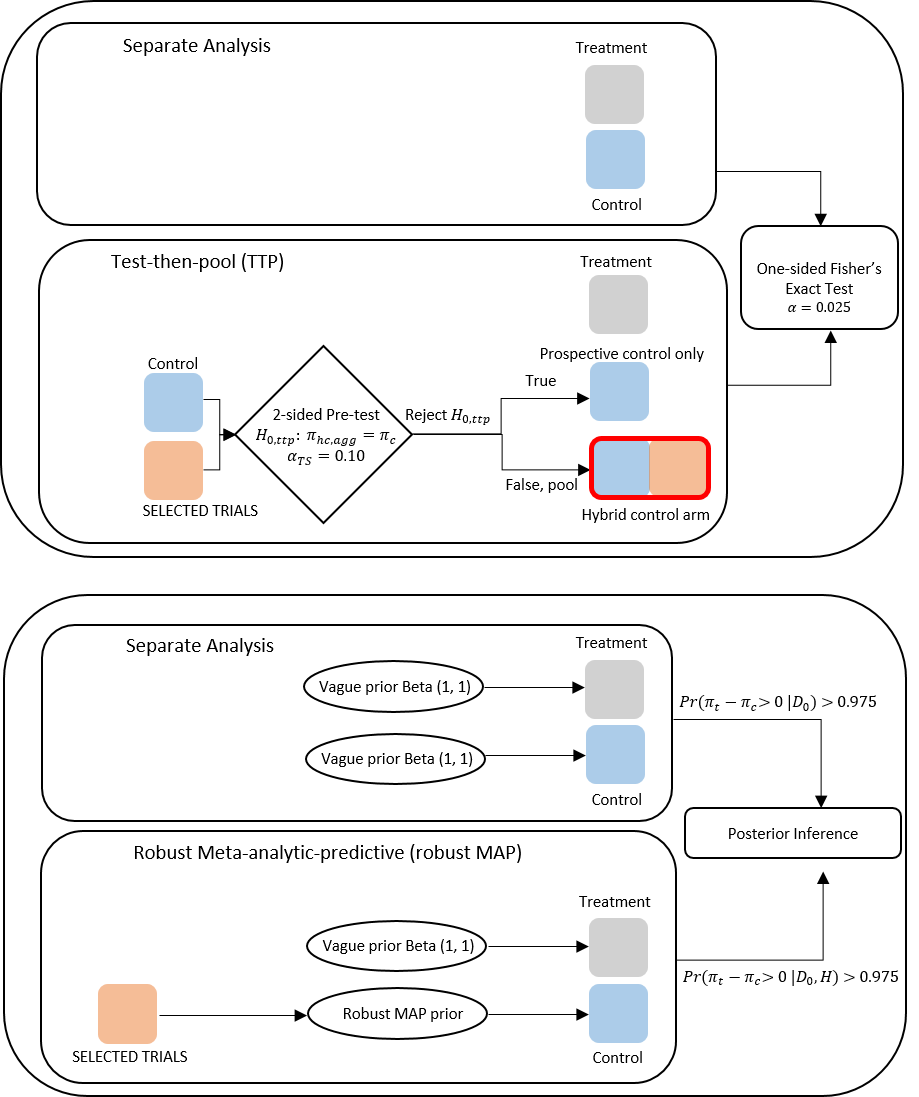}}
\caption{Analysis framework of the simulation study.
   Top: Frequentist analysis. The top panel shows separate analysis with a one-sided Fisher’s exact test using prospective trial data only (gray and blue). The bottom panel shows test-then-pool, which uses a two-sided pre-test to assess similarity between the selected historical control and the prospective control; after the pooling decision, a one-sided Fisher’s exact test is performed for treatment effect assessment.
   Bottom: Bayesian analysis. The top panel shows separate analysis fitting a vague prior Beta(1,1) to both arms. The bottom panel shows the robust MAP approach, where the robust MAP prior is obtained from selected historical trials (orange). Circles indicate the prior of the analysis, which is then updated with the prospective treatment and control data (gray and blue) to conduct posterior inference.}
  \label{fig:analysis_methods}
\end{figure}

We consider hybrid RCTs, based on a prospective trial with 1:1 randomization and a pool of \(k\) historical trials for which we only consider the control data,  with a binary endpoint. We assume that the historical controls used in the analysis are selected based on specific selection rules. The hybrid RCT then incorporates these historical controls to augment the prospective control arm in the analysis (Figure~\ref{fig:hybrid_RCT}b). As a reference, we compare these to a stand-alone RCT consisting of the prospective trial only (Figure~\ref{fig:hybrid_RCT}a). For both types of trials, we consider both frequentist and Bayesian analysis methods (Figure~\ref{fig:analysis_methods}). For the stand-alone RCT  Fisher's Exact test is used in the frequentist framework, while in the Bayesian framework, both arms are fitted with a uniform prior, (i.e. Beta(1,1) prior for dichotomous outcome). For the hybrid RCT dynamic borrowing is implemented with, a test-then-pool approach in the frequentist framework, whereas a robust MAP approach is utilized for the Bayesian analysis. Because in practice hybrid RCT often randomize more participants to the experimental treatment to maximize potential benefit, we also explore unequal randomization ratios of 2:1 and 3:1 while keeping the total sample size fixed in the prospective part of hybrid RCT. 

The parameter of interest is the treatment effect defined as the risk difference (RD):
\[
RD = \pi_t-\pi_c,
\]
where $\pi_t$ and $\pi_c$ are the true response probabilities for patients eligible for the prospective trial under treatment and control, respectively. External controls (historical trials) may inform the estimation of $\pi_c$ through pooling or prior augmentation. We test the hypotheses
\[
H_0: RD \le 0 \qquad \text{vs.} \qquad H_1: RD > 0.
\]

While there are several methods available for leveraging information of multiple historical trials, in this study we focus on TTP and robust MAP. The risk difference is estimated as the difference between the mean response rates in the treatment and control arms. For the frequentist analyses, we use sample means from the prospective trial, with the option to pool historical controls when appropriate; for the Bayesian analyses, we use posterior means under either vague prior or priors derived from historical controls.

\paragraph*{Separate Analysis}\label{separate}
A stand-alone RCT implies that only data from the prospective trial itself are used in the analysis. This is denoted by the term separate analysis which uses only prospective trial data and serves as a \emph{no-borrowing} benchmark.
As frequentist test, a one-sided Fisher's exact test at level $\alpha=0.025$ is performed.
In the corresponding Bayesian analysis we assume for both arms $\mathrm{Beta}(1,1)$ priors and declare success when
\[
\Pr\!\big(\pi_t-\pi_c>0 \,\big|\, \Dnew\big) > 0.975,
\]
$\Dnew$ denotes the prospective trial data in both arms.

\paragraph*{Test-then-pool}\label{TTP}
Historical controls from $k$ trials are first selected (Section~\ref{selection rules}), then equally weighted and pooled into a single historical control dataset. Consistency between historical and prospective controls is assessed by a two-sided pre-test (Fisher's exact test) at level $\alpha_{TS}$, testing the hypotheses
\begin{equation}
\label{eq:consistency_test}
H_{0,TTP}:\ \pi_{hc}=\pi_{c}
\quad\text{vs.}\quad
H_{1,TTP}:\ \pi_{hc}\neq\pi_{c},
%\qquad \alpha_{\mathrm{TS}}.
\end{equation}

where $\pi_{hc}$ and $\pi_c$ are the historical and prospective control response rates, respectively. Fail to reject $H_{0,TTP}$ indicates that there is no significant difference between the historical  and prospective controls. Therefore, in this case the historical control data is pooled with the prospective control data to form a hybrid control arm for the subsequent treatment-control comparison. On the other hand, if the difference between historical and prospective controls is significant at level $\alpha_{TS}$, the analysis will proceed with the prospective control only. In both cases the subsequent treatment-control comparison is performed with a one-sided  Fisher's exact test at level $\alpha=0.025$ using the pooled or the prospective controls, only.

\paragraph*{Robust Meta-analytic-predictive Priors}\label{Robust MAP}
Neuenschwander et al.\cite{neuenschwander_summarizing_2010} proposed to summarize historical information in a meta-analysis framework. A meta-analytic-predictive (MAP) prior is derived from the historical data to inform the prospective trial. This MAP prior has no simple parametric form but can be approximated by fitting a finite mixture of natural conjugate priors \cite{dalal_approximating_1983}. For example, for binary endpoints, the MAP prior derived from binomial controls is well approximated by a mixture of Beta distributions. Schmidli et al.\cite{schmidli_robust_2014} further proposed a robust version of the MAP prior that accounts for potential prior-data conflict. This robustification is achieved by  averaging a vague prior and the MAP prior, with a prespecified weight ($w_R$) that corresponds to probability of conflict between historical control and new trial, see  ~(\ref{eq:rmap_mixture}). $w_R = 1$ assigns full weight on the vague prior and leads to separate analysis, while $w_R=0$ corresponds to a full MAP prior. The resulting prior is given by

\begin{equation}
\label{eq:rmap_mixture}
\begin{aligned}
\hat{p}_{H,\mathrm{robust}}(\pi_{new}) = w_R \overbrace{\text{Beta}(1,1)}^{\text{vague prior}} + (1-w_R) \overbrace{\sum_{\ell=1}^K w_\ell \text{Beta}(\pi_{new}|a_\ell, b_\ell)}^{\text{MAP prior from historical trials}}
&\quad \text{with } \sum_{\ell=1}^K w_\ell = 1,
%\quad \text{with } w_R = 0.1
\end{aligned}
\end{equation}

where $\ell$ indexes mixture components. Historical controls were selected using the rules in Section~\ref{selection rules} prior to fitting the robust MAP. The subsequent robust MAP analysis borrows from the selected historical trials, with the degree of borrowing governed by between-trial similarity in the MAP/robust MAP framework. Treatment arm in the prospective trial is informed by vague prior Beta(1,1). We declare success if
\[
\Pr\!\big(\pi_t-\pi_c>0 \,\big|\, \Dnew,\Dhist\big) > 0.975,
\]
where $\pi_t$ and $\pi_c$ denote the treatment and control response rates, $\Dnew$ the prospective trial data, and $\Dhist$ the historical data used for borrowing. Full posterior updating and computation of posterior superiority probabilities are detailed in Supplementary Material S1. Analyses were implemented in \textsf{R} with the \texttt{RBesT}~\cite{weber_applying_2021} package.

\section{Simulation study}\label{sec:sim}
The main objective of this simulation study is to evaluate how the operating characteristics of a hybrid clinical trial design are affected by selection of historical controls. Simulations have been performed to evaluate marginal operating characteristics averaging over the prior of the control response rates and outcomes of the historical trials: type I error rate, power, bias, root mean square error (RMSE), and effective sample size (ESS) of the selection rules described in Section~\ref{selection rules}. See Table~\ref{tab:oc} for a summary of the assessed operating characteristics.

\begin{table*}[!t]%
\centering%
\small
\caption{Marginal operating characteristics estimated in the simulation study averaged over the priors and the sampling distribution. The power is evaluated under $H_{1}$ (true $RD=0.20$) and the type I error rate is evaluated under $H_{0}$ (true $RD=0$). Bias and RMSE are evaluated under both $H_{0}$ and $H_{1}$. \label{tab:oc}}%
\begin{tabular*}{\textwidth}{@{\extracolsep\fill}p{2.4cm}p{3.6cm}p{3.8cm}p{4.7cm}@{\extracolsep\fill}}
\toprule
\textbf{Measure} & \textbf{Definition} & \textbf{Estimation} & \textbf{Description} \\
\midrule
Bias & $E[\hat{RD} - RD]$ & $\displaystyle \frac{1}{n_{\text{sim}}}\sum_{r=1}^{n_{\text{sim}}}
              \bigl(\hat{RD}^{(r)}-RD\bigr)$ & Systematic deviation of the estimated treatment effect from the true value. \\
RMSE & $\sqrt{E[(\hat{RD}-RD)^2]}$ & $\displaystyle \sqrt{\frac{1}{n_{\text{sim}}}\sum_{r=1}^{n_{\text{sim}}}
              \bigl(\hat{RD}^{(r)}-RD\bigr)^2}$ & Average magnitude of estimation error, in the same units as $RD$. \\
Power & $\Pr(\text{reject } H_{0}\mid RD=0.20)$ & Proportion of rejections & Probability of rejecting $H_{0}$ when the true effect is 0.2 (evaluated under $H_{1}$). \\
Type I error rate & $\Pr(\text{reject } H_{0}\mid RD=0)$ & Proportion of false positives & Probability of rejecting $H_{0}$ when treatment has no effect (evaluated under $H_{0}$). \\
ESS & Effective sample size of the prior & Average prior ESS calculated using the expected local-information-ratio (ELIR) method \cite{neuenschwander_predictively_2020} & Quantifies historical information contributed by borrowing, expressed as ESS. Averaged over replicates; varies with the selection rule. \\
\bottomrule
\end{tabular*}
\end{table*}

 \subsection{Simulation setup}
\paragraph*{Data-generating mechanism.} 
We consider $k$ historical control arms indexed by $i=1,\dots,k$. For each historical arm with sample size $n_i$, the binary response count $Y_i$ follows a binomial distribution $Y_i \sim \mathrm{Bin}(n_i,\pi_i)$, where $\pi_i\in(0,1)$ is the event probability simulated from
\begin{equation}
  \theta_i = \mathrm{logit}(\pi_i) = \beta_{0} + u_i,
  \qquad u_i \stackrel{\mathrm{iid}}{\sim} \mathcal{N}(0,\tau^{2}).
  \label{eq:theta_logit}
\end{equation}

\noindent$\theta_i$ is the log-odds of $\pi_i$ and $u_i$ is the trial-specific random effect, which models the between-trial heterogeneity and has variance $\tau^2$. The prospective trial is indexed by $i=k+1$ and has two arms: a concurrent control arm with event probability $\pi_c$ and a treatment arm with event probability $\pi_t$. Their log-odds are $\theta_c=\mathrm{logit}(\pi_c)=\beta_{0}+u_{k+1}$ and $\theta_t=\mathrm{logit}(\pi_t)=\beta_{0}+\beta_{1}+u_{k+1}$, where $u_{k+1}\sim\mathcal{N}(0,\tau^2)$. Thus the historical controls and the concurrent control arm share the same random-effects distribution $\theta_1,\dots,\theta_k,\theta_c\stackrel{\mathrm{iid}}{\sim}\mathcal{N}(\beta_{0},\tau^{2})$.
The fixed effects $(\beta_{0},\beta_{1})$ were chosen on the logit scale to give a control response rate of $0.20$ and a planned risk difference of $0.20$ between treatment and control. Specifically,
$\beta_0 = \mathrm{logit}(0.20)=\log\!\left(\tfrac{0.20}{0.80}\right)$,
$\beta_1 = \mathrm{logit}(0.40)-\mathrm{logit}(0.20),$
so that at the reference level $u_{k+1}=0$ this yields $\pi_c=0.20$ and $\pi_t=0.40$, corresponding to a risk difference of $0.20$. Note that while the parameters of the historical and prospective control response rates are sampled from a prior, the treatment effect is assumed to be fixed.

\paragraph*{Trial design choices and scenario assumptions} 
Following simulation reporting guidelines \cite{morris_using_2019,friede_refinement_2010}, we distinguish trial design choices (D) from scenario assumptions (A). In all simulations, TTP was implemented with a two-sided Fisher’s exact pre-test at level $\alpha_{TS}=0.10$ for the pooling decision, followed by a one-sided Fisher’s exact test to assess the treatment effect. Robust MAP was implemented with $w_R=0.1$.
Table~\ref{tab:scenfac} summarizes the design options that were varied in the simulation study, such as the sample sizes of the prospective trial $n_\text{total}$ and randomization ratios. In addition, the considered assumptions (between-trial heterogeneity $\tau$, number of historical trials $k$, sample size of historical controls per trial $n_{hc}$, and the true prospective control response rate $\pi_c$) are listed. In particular, we consider the main simulation Scenarios 1a, 1b and two extensions, Scenarios 2 and 3, described below.

\begin{itemize}
\item Scenario 1: all historical controls are simulated with response rate $\pi_{hc}$ according to (\ref{eq:theta_logit}) where $E(\pi_{hc})=0.20$. 
\begin{itemize}
    \item[-] Scenario 1a \emph{Exchangeable}:
    When the prospective control response rates also satisfy $E(\pi_{c})=0.20$, the historical and prospective controls response rates ($\pi_{hc}, \pi_{c}$) are \emph{exchangeable}; their study-specific logits are i.i.d.\ $\mathcal N(\beta_0,\tau^{2})$. Thus, they are drawn from the same distribution (differences across historical and prospective controls arise only from the study-specific random effects $u_i$), and on average (averaged over the random effects) the response rates in the historical and concurrent controls are the same. 
    \item[-] Scenario 1b \emph{Prospective-historical distributional shift}:   $E(\pi_{c})\not=0.20$ such that the prospective control arm shows on average a prior-data conflict.

\end{itemize}

    \item Scenario 2 \emph{Time trend:} We impose a linear time effect on the historical controls with $\beta_{2}=-0.05$ on the logit scale (only for the case of $k=8$ historical control trials each with a sample size of $n_{hc}=30$). The drift enters as $\beta_{2}(k-i+1)$ in (\ref{eq:theta_logit}) and is anchored at zero for the prospective trial ($i=k+1$), so historical controls drift from the prospective baseline. Consequently, historical response rates rise from $\approx 0.14$ (earliest) toward $\approx 0.20$ (most recent), while the prospective control response rate matches the scenario 1 ($\approx 0.20$); thus any differences arise solely from the trend in historical studies.
    
    \item Scenario 3 \emph{Large prospective trial:} In addition to the scenario 1 ($n_\text{total}=60,180$ with $RD=0.20$), we generated larger trials ($n_\text{total}=500$) with smaller effects $RD=(0.0635,\,0.1152)$, while fixing the historical trials with sample size per arm of $n_{hc}=(30, 60)$. These values were calibrated such that the no-borrowing design achieved approximately the same power ($\approx39\%$ and $84\%$) as the smaller trials, thereby isolating the gain from borrowing.
\end{itemize}

\begin{table*}[!t]%
\centering%
\small
\caption{Factors varied in the simulation, classified as Design (D) or Assumption (A), with investigated values and description.\label{tab:scenfac}}%
\begin{tabular*}{\textwidth}{@{\extracolsep\fill}
p{3.2cm} p{1cm} p{3.6cm} p{7cm}
@{}}
\toprule
\textbf{Factor} & \textbf{Type} & \textbf{Investigated values} & \textbf{Description} \\
\midrule
$\tau$ & A & 0.10,\;0.30,\;0.50 & Between-trial heterogeneity; values motivated by Neuenschwander et al.\ (2010) \cite{neuenschwander_summarizing_2010}. \\
$k$ & A & 4,\;8 & Number of historical trials (control arm only). \\
$n_{hc}$ & A & 30,\;90 & Historical control sample size per trial. \\
$n_{\text{total}}$ & D & 60,\;180 & Prospective trial total sample size. \\
Randomization ratio & D & 1:1,\;2:1,\;3:1 & Allocation ratio of prospective trial (treatment:control), with fixed $n_\text{total}$. \\
$\pi_{c}$ & A & 0.15-0.75 (step 0.05) & True prospective control response rate. \\
\bottomrule
\end{tabular*}
\begin{tablenotes}\scriptsize
\item \textit{Type:} D = Design; A = Assumption.
\end{tablenotes}
\end{table*}

\begin{table*}[!t]%
\centering%
\small
\caption{Simulation scenarios for prospective trial. The scenario numbers are used for reference in the Results section.\label{tab:scenfamilies}}%
\begin{tabular*}{\textwidth}{@{\extracolsep\fill}lp{3.8cm}p{11cm}@{\extracolsep\fill}}
\toprule
\textbf{Scenario} & \textbf{Name} & \textbf{Description} \\
\midrule
1a & Exchangeable & No distributional shift, $E(\pi_{hc})=E(\pi_{c})=0.20$, historical and prospective control response rates $\pi_{hc}, \pi_{c}$ are exchangeable. \\
1b &Prospective-historical distributional shift & $E(\pi_{hc})=0.20$, prospective $E(\pi_{c})$ varied from 0.15 to 0.75 (step 0.05). Represents prior-data conflict. \\
2 & Time trend & Linear logit drift ($\beta_{2}=-0.05$) across historical trials for $k=8,\;n_{hc}=30$; prospective trial fixed at $E(\pi_{c})=0.20$, earliest historical trial $E(\pi_{hc})\approx 0.14$. \\
3 & Large prospective trial & Larger prospective trial with total sample size ($n_{\text{total}}$) = 500 with smaller effects $RD=(0.0635,\,0.1152)$, calibrated so that the no-borrowing design achieves $\approx39\%$ and $84\%$ power, respectively. \\
\bottomrule
\end{tabular*}
\begin{tablenotes}\scriptsize
\item $RD$ denotes risk difference in prospective trial used for calibration. $\pi_{hc}$: historical control trials response rate; $\pi_{c}$: prospective trial control response rate
\end{tablenotes}
\end{table*}

Table~\ref{tab:scenfamilies} summarizes the above simulation scenarios.
For each scenario and design $10\,000$ simulation runs were performed. All scenarios were simulated with the same random number seed to reduce the Monte Carlo error of between scenario comparisons. 
For each simulation run, we simulated both, historical and prospective data, to obtain \emph{marginal} operating characteristics, i.e.\ operating characteristics that are averaged over the variability in both the response rates and the sampling distributions of historical and prospective data, rather than conditional on a fixed set of historical controls. 

\paragraph*{Considered selection rules}\label{selection rules}
To investigate how the selection of historical controls (and discarding some of them) may impact important operating characteristics, we explore a range of different selection rules summarized in  Table~\ref{tab: Selection rules}. These selection rules are not proposed for practical use, but were chosen as illustrative examples of rules that might be applied when designing a hybrid RCT. 
The rules range from approaches unrelated to the observed effect, such as selecting all $k$ available historical controls ("Full selection") or randomly selecting some of the control arms ("Random selection"), to rules that use information on the observed response rates. Among the latter some are more extreme than others. E.g., "Drop-the-best" simply drops a single historical control, i.e., the one with the largest ("best") observed response rate. Another rule drops all historical controls with a too positive outcome, i.e., where the observed response rate is above a certain threshold.

\begin{center}
\begin{table*}[!h]%
\small
\caption{Description of selection rules when $k$ historical trials are available.\label{tab: Selection rules}}
\begin{tabular*}{\textwidth}{@{\extracolsep\fill} l p{7.6cm} c p{3.4cm} @{}}
\toprule
\textbf{Selection rule} & \textbf{Description} & \textbf{ODS$^{\mathrm{a}}$} & \textbf{Number of selected trials$^{\mathrm{b}}$} \\
\midrule
Full selection & Utilize all $k$ available historical trials as external controls ($k=4,8$). & N & $k$ \\
Random selection & Randomly drop a historical control. & N & $k-1$ \\
Drop-the-best & Drop the historical control with the highest response rate. & Y & $k-1$ \\
Threshold selection & Select all historical controls with response rates $\le$ a prespecified threshold (e.g., $20\%$; see Figure~\ref{fig:selection_rules}b). & Y & random \\
Optimal power selection & Choose the subset that maximizes conditional power, assuming known current-trial parameters. & Y & random \\
Separate analysis & Discard all historical controls (no borrowing). & -- & -- \\
\bottomrule
\end{tabular*}
\begin{tablenotes}\scriptsize
\item[$^{\mathrm{a}}$] $^{\mathrm{a}}$ODS: Outcome-dependent selection that indicates whether the observed response rates are used. Y = Yes; N = No.
\item[$^{\mathrm{b}}$] $^{\mathrm{b}}$“random” indicates that the number of selected historical trials varies by rule (e.g., threshold or optimal power selection).
\end{tablenotes}
\end{table*}
\end{center}

\begin{figure}[!htbp]
\centerline{\includegraphics[width=\textwidth]{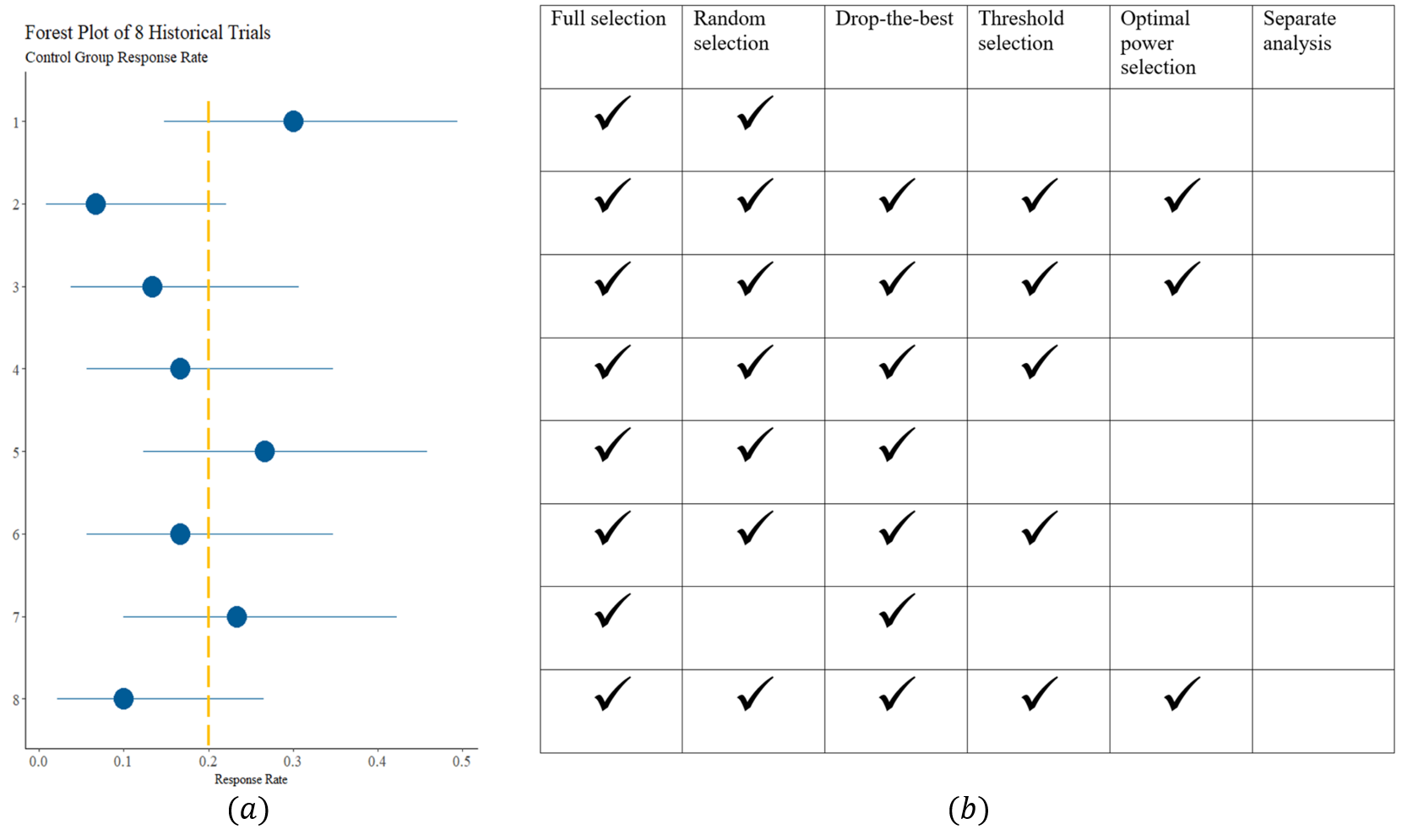}}
\caption{Illustration of different selection rules implemented in this study with $k=8$ historical trials. (a) Forest plot shows control group response rates with 95\% confidence intervals. The yellow line indicates the threshold ($\pi_{hc}=0.20$). (b) Table that indicates which historical trials are selected with indicators ($\checkmark$) under each rule. For description of the selection rules see Table \ref{tab: Selection rules}. The historical trials are numbered in chronological order.}
  \label{fig:selection_rules}
\end{figure}

Figure~\ref{fig:selection_rules} illustrates the considered selection rules based an exemplary set of $k=8$ historical controls. We define outcome-dependent selection (ODS) as any rule that uses observed outcomes from historical studies to decide which to borrow. In  Table~\ref{tab: Selection rules}, Drop-the-best, threshold selection and optimal power selection are ODS rules. 

The optimal power selection rule chooses the subset of historical controls that leads to the maximum conditional power to reject the null hypothesis in the prospective trial. Here we assume that the prospective trial has not been performed yet but the distributions from which observations of the prospective trial are sampled, are known.
The conditional power is defined as the conditional probability, given the data of the selected historical controls, that the prespecified Bayesian success criterion, $\Pr(\pi_t - \pi_c > 0 \mid \Dnew, \Dhist) > 0.975$, will be met after the prospective trial data have been observed. Here it is assumed that  the prospective trial data is sampled with known response rates  $(\pi_t^\star,\pi_{c}^\star)$ and the trial has sample sizes $(n_t,n_{c})$.  
Full technical details are provided in appendix for the optimal power selection rule. Please note that for the threshold and optimal power selection rule there might be the situation that none or only one historical trial will be selected (see online supplement S2).

\subsection{Results}\label{sec3}

\subsubsection{Scenario 1a: Historical control response rates and prospective control response rate are \emph{exchangeable}}

In the frequentist setting, TTP with full or random selection maintains the type I error rate below 0.025 only when between-trial heterogeneity is low. ODS rules inflate the type I error rate, especially with larger treatment:control randomization ratios (Supplementary Figure S1). In the Bayesian analysis using a robust MAP prior, full and random selection keep the type I error rate close to the nominal 0.025 under low to moderate heterogeneity (Supplementary Figure S1) and when historical studies are small ($n_{hc}$ = 30). Under high heterogeneity, borrowing with these selection rules exceeds the nominal type I error level. With robust MAP, type I error inflation is more severe when borrowing from smaller historical studies ($n_{hc}=30$), but this does not hold for TTP (Supplementary Figure S2). 
Across both analysis frameworks, ODS rules inflate the type I error rate, with the greatest inflation at high heterogeneity. When more historical trials are available ($k=8$), ODS rules that select a random number of studies (e.g., threshold selection, optimal power selection) have more degrees of freedom and tend to cause more severe inflation. Type I error inflation is much more severe with ODS rules when the historical study pool has large $k$ and small $n_{hc}$ than when it has small $k$ and large $n_{hc}$, regardless of the level of between-trial heterogeneity (Supplementary Figure S2). This warrants particular care when several small historical trials are available. 

ODS rules lead to a larger power than full and random selection when the prospective trial is small ($n_{hc} = 30, n_\text{total} = 60$, Supplementary Figure S4). ODS rules provide greatest power advantage when between-trial heterogeneity is large (Supplementary Figure S3). The largest relative power gain from borrowing occurs with a 3:1 randomization ratio, despite this ratio has lower absolute power than 1:1, regardless the selection rules (Supplementary Figure S3). More patients enjoy the benefit of experimental treatment arm while achieving similar power when we borrow the data. When between-trial heterogeneity is small ($\tau = 0.1$) and the historical data set is large ($n_{hc}=90$), all historical controls are highly similar to the prospective control. In this situation, ODS rules offer no power gain over simply using all or randomly chosen historical controls, because nearly every trial in the pool contributes useful information.

With TTP, ODS rules introduce positive bias in the treatment effect estimates. This bias is larger with threshold selection and optimal power selection and  when the sample size in the prospective trial is small ($n_\text{total}=60$). With this sample size, the biases for these two selections are around 2-3 percentage points. In the Bayesian framework, if a uniform prior is applied in both arms, even in the separate analysis a negative bias in the treatment effect estimate, under $H_1$ (Supplementary Figure S5) is introduced. This a property of the Bayesian analysis when the sample size is small. The bias becomes more pronounced if unequal randomization is applied and more patients are allocated to the treatment arm. ODS rules introduce a positive bias resulting in an overestimation of the treatment effect (Supplementary Figure S5-S6). Higher numbers of historical studies ($k = 8$) increases the bias introduced by ODS when the robust MAP is used (Supplementary Figure S6). Similar findings are observed under $H_0$ (Supplementary Figure S7-S8).

Borrowing historical control reduces RMSE whenever heterogeneity is low ($\tau=0.1$), regardless of the selection rules (Supplementary Figure S9). At moderate heterogeneity ($\tau=0.3$), a RMSE reduction is seen only for $n_{hc}=30$. Threshold selection and optimal power selection yield higher RMSE compared to other selection rules across all levels of between-trial heterogeneity ($\tau$), the number of historical studies ($k$), or the sample size within each historical study ($n_{hc}$)(Supplementary Figure S9-S10). Similar findings are observed under $H_0$ (Supplementary Figure S11-S12).
RMSE increases with higher randomization ratio (keeping  $n_\text{total}$ fixed) as unequal treatment:control allocation increases the variance of the RD estimate. Borrowing historical control can reduce the impact of randomization ratio when $\tau=0.1, 0.3$. 

For the robust MAP, we report the ESS, representing the amount of information that is used from the selected historical controls in the treatment control comparison,  for the different selection rules in the Supplementary Tables S1-S4. With increasing between-trial heterogeneity ($\tau$), the ESS decreases across all settings of $k, n_{hc}$ and selection rules. This ESS reduction with increasing $\tau$ is smaller with more flexible ODS rules like threshold selection and optimal power selection, because these rules keep a more homogeneous subset of studies, at the cost of selection bias. For larger $\tau$ or $k$ ODS rules can lead to higher ESS than full or random selection because a large heterogeneity or a larger number of historical studies that can be selected enables more extreme selections. For the optimal power selection, prospective trial's treatment:control randomization ratio affects the sample size $(n_t,n_{c})$ and thus influences the selection. However, changing the treatment:control ratio has only minor, non-systematic effects on ESS (Supplementary Table S3-S4). The ESS with increasing randomization ratio remains stable or decreases slightly. Both drop-the-best and random selection retain $k-1$ studies, but drop-the-best yields larger ESS because removing the highest response historical trial reduces heterogeneity, at the cost of potential bias.

\subsubsection{Scenario 1b: Distributional shift in prospective control population resulting prior-data conflict}
\begin{figure}[!htbp]
\centerline{\includegraphics[width=\textwidth]{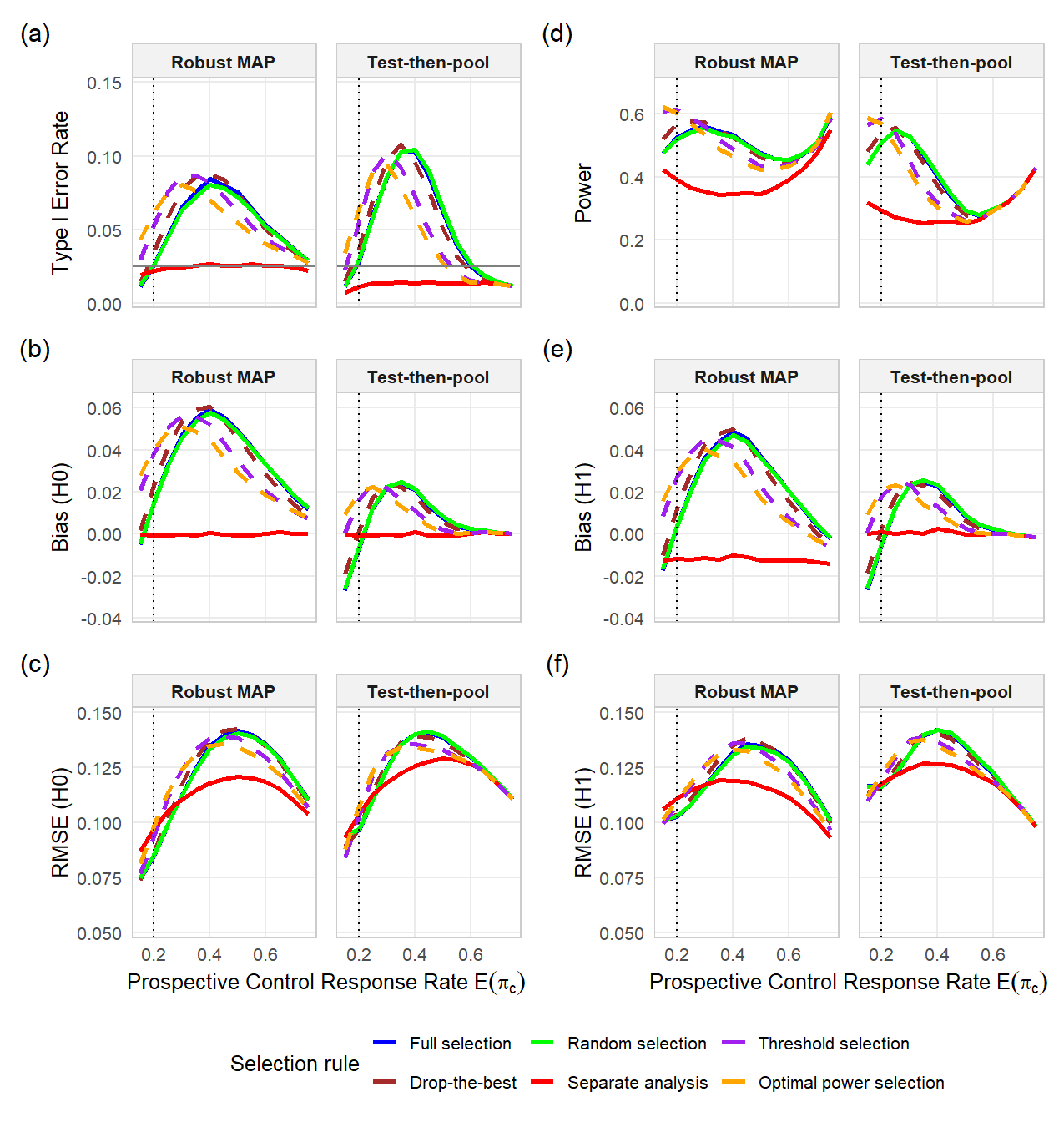}}
\caption{Marginal operating characteristics of selection rules with an initial set of $k=8$ historical control trials each with a sample size of $n_{hc}=30$ and assuming a moderate between trial heterogeneity of $\tau=0.30$, prospective trial total sample size $n_\text{total}=60$ (for more details on the simulation setup see Section~\ref{sec:sim}). (a)–(c) $H_0$: type I error rate, bias, RMSE. (d)–(f) $H_1$: power, bias, RMSE. x-axis: prospective control response rate; the vertical line at $x=0.20$ indicates exchangeability, where maximum amount of borrowing is appropriate; dashed lines denote outcome-dependent selections.}
  \label{fig: OC main}
\end{figure}

Figure~\ref{fig: OC main} illustrates the impact of \emph{prospective–historical distributional shift}, which is the case when historical data differs from the prospective control response rates (Scenario 1b). Borrowing will only be appropriate when the prospective control response rates are similar to historical control response rates (around 0.20, vertical line). In this case, full and random selection show no bias, lower RMSE, higher power, and well-controlled type I error rate compared to the separate analysis. ODS rules (dashed lines) achieve higher power but at the cost of greater bias, higher RMSE (Figure~\ref{fig: OC main}d-f), and inflated type I error rate (Figure~\ref{fig: OC main}a). When the prospective control response rate is larger than expected, we observed dynamic behavior of the borrowing methods, an initial increase followed by a decline in these operating metrics. Inflated type I error rates are expected whenever information is borrowed, consistent with the findings of Kopp-Schneider \cite{koppschneider_power_2020}. In this setting, the performance curves for ODS rules eventually intersect those of full and random selection because ODS rules preferentially draw on historical studies with lower response rates. As the prospective control response rate increases, the dynamic borrowing methods then borrow less information with ODS rules as the conflict grows and revert to no borrowing earlier than full and random selection. These findings hold across all simulation scenarios we examined, regardless of the values of $\tau, k, n_{hc}$, although the magnitude of the effects varies.

\subsubsection{Scenarios 2 and 3: Time trend and large prospective trial scenarios}
A time trend was applied by adding a linear drift on the logit scale $\beta_{1}=-0.05$ to all historical trials (Scenario 2). The resulting mismatch of historical and prospective controls increased bias, root mean-squared error, and the one-sided type I error rate for every selection rules (Table~\ref{tab:oc_scenarios_bayes}); even full and random selection, which do not favor extreme studies, now lead to an inflation of the nominal type I error rate. The ODS rules amplify the problem: drop-the-best, threshold, and optimal power selection show larger increases in bias and type I error rate while providing only modest power gains. When temporal drift is present, borrowing therefore requires a careful assessment of comparability and strong justification.

In the large prospective trial scenario (Scenario 3) we set the the true risk difference to 0.0635 and increased each trial arm to 250 participants. Under this design, bias and RMSE decrease for all methods as the prospective trial data dominate the posterior. However, the three ODS rules still show larger biases and type I error rates inflation, although both distortions are smaller than in Scenario 1. Power differences are also smaller, because the larger prospective control arm leaves comparatively little room for historical data augmentation.

Sensitivity analyses with a more relaxed prior for $\tau$ have been conducted (as the results are similar, data are not shown).

\begin{center}
\small
\begin{table*}[!h]%
\setlength\tabcolsep{2.5pt}
\caption{Marginal operating characteristics (\%) with an initial set of $k=8$ historical control trials each with a sample size of $n_{hc}=30$ and assuming a moderate between trial heterogeneity of $\tau=0.30$ (for more details on the simulation setup see Section~\ref{sec:sim}).\label{tab:oc_scenarios_bayes}}
\begin{tabular*}{\textwidth}{@{\extracolsep\fill} l cccc cccc cccc @{}}
\toprule
& \multicolumn{4}{@{}c}{\textbf{Exchangeable}$^{\mathrm{a}}$} 
& \multicolumn{4}{@{}c}{\textbf{Time-trend}$^{\mathrm{b}}$} 
& \multicolumn{4}{@{}c}{\textbf{Large prospective trial}$^{\mathrm{c}}$} \\
\cmidrule(lr){2-5}\cmidrule(lr){6-9}\cmidrule(lr){10-13}
\textbf{Selection rules} 
& \textbf{T1E} & \textbf{Power} & \textbf{Bias} & \textbf{RMSE}
& \textbf{T1E} & \textbf{Power} & \textbf{Bias} & \textbf{RMSE}
& \textbf{T1E} & \textbf{Power} & \textbf{Bias} & \textbf{RMSE} \\
\midrule
Separate analysis          &  2.20 & 39.05 & $-1.17$ & 11.09  & 2.20 & 39.05 & $-1.17$ & 11.09  & 2.57 & 38.87 & $-0.04$ & 3.85 \\
Full selection             &  2.63 & 52.75 &  0.38   & 10.28  & 3.95 & 58.32 &  1.85   & 10.52  & 2.83 & 42.91 &  0.18   & 3.90 \\
Random selection           &  2.60 & 52.21 &  0.37   & 10.31  & 3.95 & 57.70 &  1.78   & 10.52  & 2.79 & 42.62 &  0.17   & 3.89 \\
Drop-the-best              &  3.48 & 56.80 &  1.21   & 10.36  & 5.27 & 61.92 &  2.62   & 10.73  & 3.28 & 45.98 &  0.41   & 3.92 \\
Threshold selection        &  5.26 & 61.43 &  2.73   & 10.76  & 6.54 & 63.96 &  3.38   & 11.01  & 4.29 & 49.38 &  0.80   & 3.95 \\
Optimal power selection    &  6.18 & 60.33 &  2.98   & 11.02  & 7.33 & 60.67 &  3.47   & 11.34  & 4.35 & 49.03 &  0.79   & 3.95 \\
\bottomrule
\end{tabular*}
\begin{tablenotes}\scriptsize
\item[$^{\rm a}$] $^{\mathrm{a}}$Exchangeable: true risk difference $RD=0.20$.
\item[$^{\rm b}$] $^{\mathrm{b}}$Time-trend: linear drift in the historical data.
\item[$^{\rm c}$] $^{\mathrm{c}}$Large prospective trial: prospective trial with $RD=0.0635$ with $n=250$ per arm.
\item $RD$: treatment effect measure, risk difference between the treatment and control arm.\\ All values are \textbf{percentages}. T1E: type I error rate at nominal one-sided level $2.5\%$. Bias: percentage-point deviation of posterior mean RD from the truth, under $H_1$, calculated by $\displaystyle \frac{1}{n_{\text{sim}}}\sum_{r=1}^{n_{\text{sim}}} \bigl(\hat{RD}^{(r)}-RD\bigr)\times 100$. RMSE: root mean square error, under $H_1$, calculated by $\displaystyle \sqrt{\frac{1}{n_{\text{sim}}}\sum_{r=1}^{n_{\text{sim}}}
              \bigl(\hat{RD}^{(r)}-RD\bigr)^2}\times 100$.
\end{tablenotes}
\end{table*}
\end{center}

\section{Case study}

We illustrate the impact of selection rules with a case study based on the ankylosing spondylitis (AS) trial data originally from Baeten et al. \cite{baeten_anti-interleukin-17a_2013} The original trial was designed as a 2-arm trial testing secukinumab against placebo with a total sample size of 30 and a 4:1 randomization ratio. The primary efficacy endpoint was the Assessment of SpondyloArthritis international Society criteria for 20\% improvement (ASAS20) at week 6. Placebo data from eight previous trials of ankylosing spondylitis were included in this trial to augment the control arm \cite{mcleod_adalimumab_2007}. This trial assumed true ASAS20 response rates of 25\% on placebo and 60\% on secukinumab, and these values were used for trial design evaluation ($\pi_{placebo}=0.25, \pi_{secukinumab}=0.60$ to select optimal power subset, RD for power evaluation is 0.35). As the original trial borrowed historical placebo data via a MAP prior, our case study assessment focus on the Bayesian framework, using the robust MAP.

\subsection{Design of the hybrid RCT and conditional operating characteristics}
\begin{figure}[!htbp]
\centerline{\includegraphics[width=\textwidth]{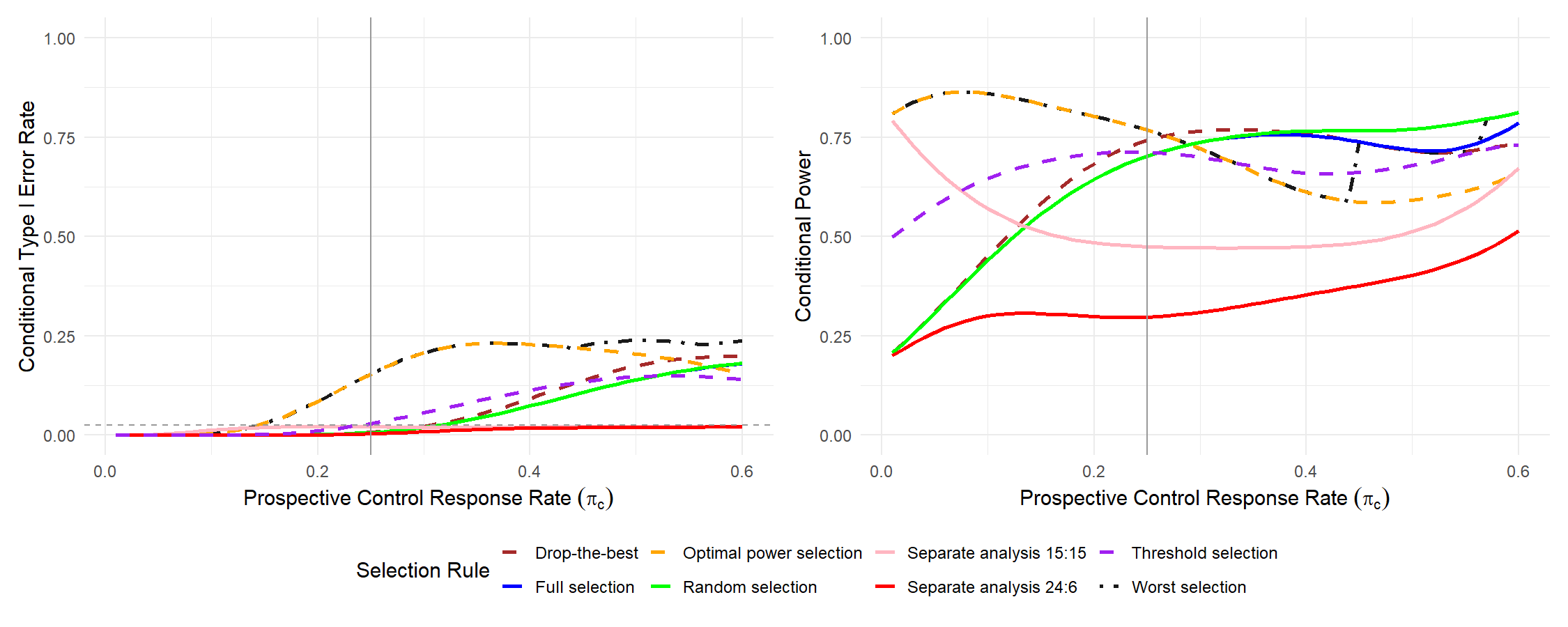}}
\caption{Conditional operating characteristics of different selection rules, based on a case study AS trial, at trial design stage. Left: Conditional type I error rates, gray dashed line is the nominal type I error rate 0.025. Right: Conditional power. The vertical line at $x = 0.25$ indicates exchangeability, where the prospective trial control response rate is similar to the mean response rate of MAP prior derived from all historical studies. x-axis: prospective control response rate; The black dashed line indicates selection that maximizes the conditional type I error rates, which is the possible worst selection in this case study.}
  \label{fig: OC case study}
\end{figure}

Supplementary Figure S15 lists the historical controls that are selected in this case study with the considered selection rules. Robust MAP priors derived from these selections were incorporated into the analysis as described in Section~\ref{Robust MAP}. At the planning stage of the prospective trial we can evaluate the trial design by calculating the conditional operating characteristics given the actual historical control dataset used and assuming a range of plausible response rates for the prospective part of the hybrid RCT. The trial designs were evaluated in terms of the conditional type I error rate and conditional power (Figure~\ref{fig: OC case study}). Conditional type I error rates are calculated under null hypothesis of $\pi_{secukinumab}=\pi_{placebo}$, conditional power evaluation is based on  $\pi_{secukinumab}=\pi_{placebo} + 0.35$. The calculation of conditional type I error rate and conditional power is an exact calculation, when plugging in range of $\pi_{c} \in [0.01, 0.60]$. To show how results change with a larger prospective sample, we repeated the evaluation for total sample sizes $n_\text{total} = 60, 300, 3000$ with same randomization ratio 4:1; Full results of these trial design evaluations are provided in Supplementary Figure S16.

 Threshold selection with threshold $\le$ 25\%  and optimal power selection inflated the conditional type I error rate even when the prospective control response matched the historical mean. The conditional type I error rates decreased in cases with larger sample size in prospective trial, but outcome-dependent selections preserved the risk of type I error rates inflation. In contrast, full selection and random selection kept the type I error rates below nominal value ($\alpha=0.025$). Gains in power from the ODS rules were modest (1$\sim$6 percentage points) compare with full and random selection when the control rate matched expectations and disappeared once the prospective control conflicted with the historical data. For $n_\text{total}=300, 3000$ we report only conditional type I error rates, because conditional power is essentially one for all methods under the alternative $\pi_{secukinumab}=\pi_{placebo} + 0.35$.

In addition to conditional type I error rate and conditional power, we also evaluated the probability of success (PoS, also known as assurance in Bayesian design literature) at the design stage. PoS averages conditional power over the joint prior distributions of treatment and control response rates, thereby reflecting the unconditional chance that the trial will declare success given the current knowledge. Table~\ref{tab:case_study} shows that PoS values differ across selection rules: ODS rules such as threshold and optimal power selection yield higher PoS (0.484 and 0.619, respectively) compared with full and random selection (0.436 and 0.432). Separate analysis results in a substantially lower PoS (0.189), indicating the additional assurance that can be obtained when borrowing external control data. Further details for the calculation of the conditional type I error rate, conditional power and PoS are available in Supplementary Material S4.

\subsection{Results of the hybrid RCT}

The results in Table~\ref{tab:case_study} are based on an intention-to-treat analysis using the Bayesian robust MAP prior approach. 14 of 24 patients responded in the secukinumab arm and 1 of 6 responded in the placebo arm. The separate analysis fitted a uniform Beta(1, 1) prior to both arms. The separate analysis failed to demonstrate superiority of the secukinumab arm over the placebo arm, with a posterior probability of treatment benefit of 0.959, see Table 6. Similarly, the frequentist separate analysis failed to yield statistical significance using a one-sided Fisher´s exact test at level $\alpha=0.025$ (see Supplementary Table S6).

\begin{center}
\begin{table*}[!h]%
\caption{Results of the Bayesian robust MAP prior analysis when using different subsets of the available historical control data depending on the selection rule illustrated for the case study ankylosing spondylitis (AS) trial data. More details on which historical controls trials have been selected for which selection rule is given in Supplementary Figure S15 in the online supplement. \label{tab:case_study}}
\begin{tabular*}{\textwidth}{@{\extracolsep\fill} l c c c c p{4cm} @{}}
\toprule
\textbf{Selection rule} & \textbf{PoS$^{\mathrm{a}}$} & \textbf{Estimate} & \textbf{95\% CI} & \textbf{$\Pr(\pi_t-\pi_c>0 \mid \Dnew, H)$} & \textbf{Number of selected trials}\\
\midrule
Full selection           & 0.436 & 0.335 & (0.090, 0.561) & 0.994 & 8\\
Random selection         & 0.432 & 0.332 & (0.076, 0.566) & 0.993 & 7\\
Drop-the-best            & 0.455 & 0.348 & (0.110, 0.566) & 0.996 & 7\\
Threshold selection      & 0.485 & 0.387 & (0.123, 0.611) & 0.995 & 3\\
Optimal power selection  & 0.619 & 0.444 & (0.223, 0.640) & 0.998 & 1\\
Separate analysis        & 0.189 & 0.327 & (-0.047, 0.624) & 0.959 & 0\\
\bottomrule
\end{tabular*}
\begin{tablenotes}\scriptsize
\item[$^{\rm a}$] PoS: Probability of success at the design stage. 
\item Estimate: Posterior treatment effect estimate (difference of posterior means between the two arms). 95\% CI: 95\% credible interval. $\Pr(\pi_t-\pi_c>0 \mid \Dnew, H)$: Posterior probability of treatment benefit, the trial is considered successful if this probability exceeds the threshold of 0.975.
\end{tablenotes}
\end{table*}
\end{center}

When using all available historical controls ("Full selection"), one can demonstrate superiority of the secukinumab arm to the placebo arm as the posterior probability of treatment effect is 0.994 ($>$0.975). The treatment effect estimate was quite similar to the separate analysis, but by using historical control data the width of the credible interval became narrower. A similar effect was observed for the random selection rule. The ODS rules still resulted in much narrower credible intervals despite leaving out some of the historical trials. However, the treatment effect estimates are much higher by leaving out historical controls where the observed effect was "too" good. The "optimal power selection" even overestimates the treatment effect by about 12 percentage points compared to the separate or full selection rule. 

However, these apparent gains for the ODS rules in terms of higher posterior probabilities of a positive treatment effect come with the previously noted inflation in the conditional type I error rate and thus emphasize the trade-off between efficiency and type I error rate control illustrated in our simulation study. Similar results are obtained with a frequentist approach when using the TTP (Supplementary Table S6).

\section{Discussion}
This study examined how the selection of external controls affects inference in hybrid randomized controlled trials. When historical and concurrent controls are comparable, borrowing improves efficiency; however, the way how controls are selected has a substantial impact on the validity of the hybrid RCT. ODS rules consistently increase bias and inflate the type I error rate for modest gains in power, whereas full and random selection provide a more conservative balance. Power gains from ODS rules are modest because ODS makes historical controls more likely to differ from prospective controls, leading to their down-weighting in the analysis. The bias patterns are robust across varying between-trial heterogeneity level $\tau$, the number and size of historical trials $k, n_{hc}$, and sensitivity analyses using a more relaxed prior for $\tau$ in the robust MAP method. In the simulations, outcomes for both the historical studies and hybrid RCTs were simulated under a range of assumptions. The corresponding operating characteristics, also referred to as marginal OCs, were computed by averaging over the sampling distribution of the historical control datasets, reflecting the long-run frequency properties of the hybrid RCT design.

Additionally, we provided conditional operating characteristics based on observed historical control datasets in the ankylosing spondylitis case study, showing the conditional type I error rates and conditional power of the hybrid trial planned with the already observed historical data.
The conditional operating characteristics show the same qualitative behavior as the marginal operating characteristics: ODS rules yield larger estimated treatment effects and higher posterior probabilities of benefit, at the price of increased type I error rates. Interestingly, the impact of ODS rules on the conditional type I error rate, persists even when the prospective trial has a large sample size of $n=3000$ (Supplementary Figure S16). This is consistent with our finding for the marginal operating characteristics in the large prospective trial scenario (Supplementary Figure S13-S14). The impact of ODS rules is not resolved by the larger sample size in the prospective trial.

In the scenario with a linear time-trend in the log-odds of response rates across the historical and the prospective trials, all borrowing strategies resulted in inflated type I error rates. ODS rules caused a further increase. This highlights the need for checks for temporal drift in the historical pool. If temporal drifts in the control response are observed, the inclusion of historical controls may be questionable. A possibility to mitigate biases induced by time trends that cannot be explained by covariate shifts, is to include calendar time as a factor in the borrowing model.

The bias caused by ODS rules decreases with less flexible selection rules. For example, for monotone selection - where all preceding historical trials are also excluded if a later trial is excluded - has a lower bias. This is illustrated for the optimal power selected historical data in the case of $k=8, n_{hc} = 30$ (see Supplementary Figure S13-S14). Using such more restrictive selection considering the temporal order of the historical trials will generally lead to less type 1 error inflation than the other non-monotone ODS rules, but overall less historical trials will be used (Supplementary Table S5).

This work has limitations. Our implementation of TTP uses a pooled pre-test in \eqref{eq:consistency_test}. An alternative TTP approach is to test each historical control separately against the prospective control (with/without multiplicity adjustment). We implemented the robust MAP with a fixed robustness weight and one family of priors (half-Normal) for the heterogeneity parameter $\tau$; other robustness weights, other MAP methods such as the modified robust MAP \cite{zhao_modified_2024}, and alternative priors for $\tau$ could have been used \cite{rover_weakly_2021}. Furthermore, we focused on binary endpoints. However, ODS will have a similar impact for time-to-event or continuous outcomes. 

The statistical methods proposed for hybrid designs are valid under the assumption that the historical controls and the prospective controls are exchangeable. However, under ODS, even if exchangeability holds for the initial pool of historical controls, this assumption is violated for the selected trials. Because the selection of controls typically occurs, after these trials have been performed, it may be impossible to demonstrate that the selection was not influenced by outcome data. Historical control data is often already publicly available in reports, publications, or registries. Prior knowledge of historical data may influence the selection process, either consciously or unconsciously, introducing a selection bias. Even if only data on a different, but correlated endpoint is available and has an impact on selection, bias will be introduced. Both, Bayesian and frequentist testing procedures are affected by such ODS. While in this work we focused on frequentist operating characteristics, the outcome-dependent selection of historical controls also invalidates Bayesian analyses. Outcome-dependent selection changes the likelihood function, and therefore Bayesian posterior computations that do not account for this will be flawed. Selecting historical trials based on their outcomes induces data-dependent priors, which compromises the validity of posterior probabilities and other Bayesian metrics such as the average type I error rate \cite{best_beyond_2025}. Therefore, under outcome-dependent selection also Bayesian inference will be invalid.

It is well known that neither Bayesian nor frequentist two-step tests can strictly control the frequentist type I error rate when incorporating historical data in the decision making \cite{koppschneider_power_2020, best_beyond_2025} if the distribution of the historical controls differs from the concurrent controls. This is a separate source of bias that comes in addition. Generally, having a higher risk under certain scenarios might be justified, e.g., in clinical trials with limited sample size such as in rare diseases. We show that if there is a selection, the operating characteristics usually presented will be biased, e.g., underestimating the actual risk of a false positive decision.

This study shows that while hybrid RCTs can improve the efficiency of stand-alone, underpowered RCT, the prespecification of external data sources remains an unresolved issue. Detailed documentation and justification of the selection process are needed, and transparency of the entire process is crucial. All potential external controls should be considered, and the selection process should be carefully assessed and reported \cite{ghadessi_roadmap_2020}. If historical trials are excluded, additional sensitivity analyses should be conducted. For example, one could think of additional analysis and simulating one additional virtual historical control study and perform a type of tipping-point analysis on how the results of a potentially left out historical trial would have to look like to change the initial result. This might help to quantify the potential selection bias. Further research is warranted.

\section*{Author contributions}
FK and MP conceived this research. All authors contributed to the planning and design of the simulation study. HCC wrote the first draft of the manuscript under the supervision of MP and FK. HCC performed the simulations and statistical analyses. FK and MP revised the manuscript. All authors discussed the results, provided comments, reviewed and approved the manuscript.

\section*{Acknowledgments}
This work is part of the SHARE-CTD Doctoral Network, which has received funding from the European Union's Research and Innovation Program Horizon Europe under the grant agreement No 101120360. For more information on the project see \href{https://www.sharectd.eu/}{www.sharectd.eu/}.

\section*{Financial disclosure}

None reported.

\section*{Conflict of interest}

The authors declare no potential conflict of interests.

\appendix

\section{Optimal Power Selection\label{app1}}%
\vspace*{12pt}

To select the historical subset that maximizes conditional power at the planning effect,
we evaluate all $2^k$ subsets $\mathcal S\subseteq\mathcal H$. 
For computational feasibility in this selection step we use a pooled informative prior
for the control arm (not the robust MAP); the robust MAP is only used later in the case study’s
trial design evaluation.

\medskip
\noindent\textbf{Setup.}
Let $\mathcal H=\{H_1,\dots,H_k\}$ be the pool of historical control trials.
For any subset $\mathcal S\subseteq\mathcal H$, let $x(\mathcal S)$ and $n_H(\mathcal S)$ be the total number
of responders and total sample size aggregated over $\mathcal S$. We take
\[
  \pi_c\mid\mathcal S \sim \mathrm{Beta}\!\bigl(x(\mathcal S),\,n_H(\mathcal S)-x(\mathcal S)\bigr),
  \qquad
  \pi_t\sim\mathrm{Beta}(1,1).
\]
For a planned two–arm trial we denote responders' counts by $(y_t,y_c)$ and sample sizes by $(n_t,n_c)$, $\Dnew=(y_t,y_c;\,n_t,n_c)$ as the notation of prospective trial data.

\medskip
\noindent\textbf{Posterior success rule and decision boundary.}
With one–sided success threshold $\gamma$ (here $\gamma=0.975$), declare success when
\(
  \Pr(\pi_t>\pi_c\mid \Dnew, \mathcal S)>\gamma.
\)
Let $P_{y_t,y_c}^{\mathcal S}=\Pr(\pi_t>\pi_c\mid \Dnew, \mathcal S)$ under the priors above. 
By Beta–Binomial conjugacy, 
\[
  \pi_t\mid y_t \sim \mathrm{Beta}(\alpha_1,\beta_1),
  \quad
  \pi_c\mid y_c,\mathcal S \sim \mathrm{Beta}(\alpha_2,\beta_2),
\]
with
\[
  \alpha_1=1+y_t,\;\; \beta_1=1+n_t-y_t,\qquad
  \alpha_2=x(\mathcal S)+y_c,\;\; \beta_2=n_H(\mathcal S)-x(\mathcal S)+n_c-y_c.
\]
Then, for $\alpha_1\in\mathbb{N}$, the posterior superiority probability admits the series
(T. Pham-Gia et al. \cite{pham-gia_bayesian_1993}):
\[
  P_{y_t,y_c}^{\mathcal S}
  \;=\;
  \sum_{k=0}^{\alpha_1-1}
  \frac{B(\alpha_2+k,\;\beta_1+\beta_2)}
       {(\beta_1+k)B(\alpha_2,\beta_2)\,B(k+1,\beta_1)}.
\]
Define the boundary as the largest treatment count not crossing $\gamma$,
\begin{equation}
d_1^{\mathcal S}(y_c)=\max\{\,y_t\le n_t:\; P_{y_t,y_c}^{\mathcal S}\le \gamma\,\}.
\label{eqS:boundary_subset}
\end{equation}
so success occurs if $y_t>d_1^{\mathcal S}(y_c)$.

\medskip
\noindent\textbf{Exact conditional power for subset selection.}
At planning values $(\pi_t^\star,\pi_c^\star)=(0.4,0.2)$,
\begin{equation}
\label{eq:power_S_final}
\mathrm{Power}(\mathcal S)=
\sum_{y_c=0}^{n_c}\binom{n_C}{y_c}(0.2)^{y_c}(0.8)^{\,n_c-y_c}
\sum_{y_t=d_1^{\mathcal S}(y_c)+1}^{n_t}\binom{n_t}{y_t}(0.4)^{y_t}(0.6)^{\,n_t-y_t}
\end{equation}
(note the inner sum starts at $d_1^{\mathcal S}(y_c)+1$ since $d_1^{\mathcal S}$ is the \emph{largest failing} $y_t$).

\medskip
\noindent\textbf{Optimal subset.}
\[
  \mathcal S^\star
  \;=\;
  \argmax_{\mathcal S\subseteq\mathcal H}\;
  \mathrm{Power}(\mathcal S),
\]
i.e.\ the historical combination that maximizes conditional power at $(0.4,0.2)$ under the pooled-Beta informative control prior.

\newpage

\section*{Supplementary Material for: Selection Bias in Hybrid Randomized Controlled Trials Using External Controls: A Simulation Study}

% Toggle + redirection of ToC writes after \beginsupplement
\newif\ifsupp
\makeatletter
\let\orig@addcontentsline\addcontentsline
\renewcommand{\addcontentsline}[3]{%
  \ifsupp
    % During supplement: redirect any ToC writes to sup.toc
    \ifnum\pdfstrcmp{#1}{toc}=0
      \orig@addcontentsline{sup}{#2}{#3}%
    \else
      \orig@addcontentsline{#1}{#2}{#3}%
    \fi
  \else
    % Outside supplement: normal behavior
    \orig@addcontentsline{#1}{#2}{#3}%
  \fi
}
% Print Supplementary ToC (sections only)
\newcommand{\suptableofcontents}{%
  \begingroup\setcounter{tocdepth}{1}\@starttoc{sup}\endgroup
}
\makeatother

\newcommand{\beginsupplement}{%
  \setcounter{section}{0}%
  \setcounter{subsection}{0}%
  \setcounter{figure}{0}%
  \setcounter{table}{0}%
  \setcounter{equation}{0}%
  \renewcommand{\thesection}{S\arabic{section}}%
  \renewcommand{\thesubsection}{S\arabic{section}.\arabic{subsection}}%
  \renewcommand{\thefigure}{S\arabic{figure}}%
  \renewcommand{\thetable}{S\arabic{table}}%
  \renewcommand{\theequation}{S\arabic{equation}}%  <-- no section coupling
  \renewcommand{\tablename}{Supplementary Table}%
  \renewcommand{\figurename}{Supplementary Figure}%
  \supptrue
}
\beginsupplement
\section*{Contents}
\suptableofcontents

\section*{Overview}

This supplementary material is organized as follows:
\begin{enumerate}[label=S\arabic*]
\item \textbf{Robust Meta-Analytic-Predictive (MAP) prior and posterior probability of superiority (binary endpoint)} details the derivation of the robust MAP prior (main manuscript Section 2) from the historical controls, the subsequent posterior update, and the computation of posterior superiority probabilities.
\item \textbf{Simulation study} provides implementation details of the simulation, along with figures and tables supporting the results presented in Section 3.2 of the main manuscript. This section also includes supporting information on monotone selection discussed in Section 5 of the main manuscript.
\item \textbf{Additional information for the case study} presents further details for the case study in Section 4 of the main manuscript, including the trials selected after applying the selection rules, conditional operating characteristics under different prospective trial settings, and the analysis results of applying various selection rules to the AS trial using TTP.
\item \textbf{Design-stage quantities (binary endpoint; Beta(1,1) for treatment; robust MAP for control)} describes the calculation of the conditional operating characteristics and probability of success in Section 4.1 of the main manuscript.
\end{enumerate}
\newpage 

%----------------------------------------------------------------------
\section{Robust Meta-Analytic-Predictive (MAP) prior and posterior probability of superiority (binary endpoint)}
\label{sup:robust_map}
%----------------------------------------------------------------------
\paragraph{Notation.}
For each study arm let $\pi\in(0,1)$ denote the response probability and $\theta=\mathrm{logit}(\pi)$ the log-odds.
Historical control arms are indexed by $i=1,\dots,k$ with random counts
$Y_i\sim\mathrm{Bin}(n_i,\pi_i)$ and $\theta_i=\mathrm{logit}(\pi_i)$.
In the prospective trial we write $\Dnew=(y_t,y_c;\,n_t,n_c)$ for the observed treatment/control counts and sample sizes.

\medskip
\noindent\textbf{Meta-Analytic-Predictive (MAP) prior on the probability scale.}
We fit a normal hierarchical model on the log-odds, with normal prior for the mean $\mu$ of log-odds and half-normal prior for the between-trial heterogeneity $\tau$
\[
  \theta_i \mid \mu,\tau^2 \sim \mathcal N(\mu,\tau^2),\qquad
  \mu\sim\mathcal N(0,2),\quad \tau\sim \mathrm{Half\!-\!Normal}(1).
\]
Given the historical data $\{\theta_i\}$ the posterior predictive for a new study on the log-odds is
\[
  p(\theta_{\text{new}}\mid \text{hist})
  = \iint \mathcal N(\theta_{\text{new}};\mu,\tau^2)\, p(\mu,\tau\mid \text{hist})\,d\mu\,d\tau.
\]
Mapping to the probability scale via $\pi_{\text{new}}=\mathrm{logit}^{-1}(\theta_{\text{new}})$ and
approximating by a finite Beta mixture (Dalal \& Hall \cite{dalal_approximating_1983}) yields
\begin{equation}
\label{eqS:map_mixture}
\widehat p_H(\pi_{\text{new}})
= \sum_{\ell=1}^{K} w_\ell\,\mathrm{Beta}(\pi_{\text{new}} \mid a_\ell,b_\ell),
\qquad w_\ell>0,\ \sum_{\ell=1}^K w_\ell=1.
\end{equation}
(Here $\ell=1,\dots,K$ indexes mixture components and is unrelated to the number of historical trials $k$.)

\medskip
\noindent\textbf{Robustification.}
To account for potential prior–data conflict, we mix the MAP prior with a vague component:
\begin{equation}
\begin{aligned}
\widehat p_{H,\mathrm{robust}}(\pi_{\text{new}})
&= w_R\,\mathrm{Beta}(\pi_{\text{new}}\mid 1,1)
  + (1-w_R)\sum_{\ell=1}^{K} w_\ell\,\mathrm{Beta}(\pi_{\text{new}}\mid a_\ell,b_\ell),\\
&\qquad w_R\in(0,1).
\end{aligned}
\end{equation}
(with $w_R=0.1$ in our simulations). In what follows we identify the new-study control with the prospective RCT control, i.e.\ $\pi_{\text{new}}\equiv \pi_c$.

\medskip
\noindent\textbf{Posterior update for the prospective control.}
Let the prospective control observe $y_c$ of $n_c$ events. By Beta–Binomial conjugacy,
each Beta component updates to a Beta posterior and the mixture weights are updated by Bayes factors.
Writing $m_\star(y_c)$ for the component-wise marginal likelihood,
\[
  m_\star(y_c)
  \;=\; \binom{n_c}{y_c}\,\frac{\mathrm{Beta}(a_\star+y_c,\;b_\star+n_c-y_c)}{\mathrm{Beta}(a_\star,b_\star)}\!,
  \qquad \star\in\{0,\;1,\dots,K\},
\]
with $(a_0,b_0)=(1,1)$. (The binomial coefficient cancels in all weight ratios.)

The posterior for $\pi_c$ is the finite mixture
\[
  \pi_c \mid y_c \;\sim\;
     \widetilde w_R\,\mathrm{Beta}(1+y_c,\,1+n_c-y_c)
     \;+\; (1-\widetilde w_R)\sum_{\ell=1}^{K}\widetilde w_\ell\,
            \mathrm{Beta}(a_\ell+y_c,\,b_\ell+n_c-y_c),
\]
with updated weights
\[
  \widetilde w_R \;=\;
  \frac{w_R\,m_0(y_c)}{w_R\,m_0(y_c) + (1-w_R)\sum_{\ell=1}^{K} w_\ell\,m_\ell(y_c)},
  \qquad
  \widetilde w_\ell \;=\; \frac{w_\ell\,m_\ell(y_c)}{\sum_{j=1}^{K} w_j\,m_j(y_c)}.
\]

\medskip
\noindent\textbf{Treatment arm.}
We use an independent Beta prior $\pi_t\sim\mathrm{Beta}(a_{t0},b_{t0})$ (e.g.\ $\mathrm{Beta}(1,1)$),
with current treatment data $y_t\sim\mathrm{Bin}(n_t,\pi_t)$. Then
\[
  \pi_t \mid y_t \sim \mathrm{Beta}(a_{t0}+y_t,\; b_{t0}+n_t-y_t).
\]

\medskip
\noindent\textbf{Joint posterior in the prospective RCT.}
With independent priors across arms, the joint posterior factorizes as
\begin{equation}
\begin{split}
p(\pi_t,\pi_c \mid y_t,y_c)
&\propto \mathrm{Bin}(y_t\mid n_t,\pi_t)\,
          \mathrm{Bin}(y_c\mid n_c,\pi_c)\,
          \mathrm{Beta}(\pi_t\mid a_{t0},b_{t0}) \\[2pt]
&\quad \times \Big[
  w_R\,\mathrm{Beta}(\pi_c\mid 1,1)
  + (1-w_R)\sum_{\ell=1}^{K} w_\ell\,\mathrm{Beta}(\pi_c\mid a_\ell,b_\ell)
\Big],
\end{split}
\label{eqS:jointpost}
\end{equation}
which yields
\[
\pi_t \mid y_t \sim \mathrm{Beta}(a_{t0}+y_t,\, b_{t0}+n_t-y_t),\\
\]
and 
\[
\pi_c \mid y_c \sim \widetilde w_R\,\mathrm{Beta}(1+y_c,\,1+n_c-y_c)
 + (1-\widetilde w_R)\sum_{\ell=1}^{K}\widetilde w_\ell\,\mathrm{Beta}(a_\ell+y_c,\, b_\ell+n_c-y_c), 
\]

with the updated weights $\widetilde w_R,\widetilde w_\ell$ defined via the component-wise marginal likelihoods $m_\star(y_c)$ as in the previous paragraph.

\medskip
\noindent\textbf{Posterior probability of superiority.}
Define success as $\pi_t>\pi_c$ on the probability scale. By independence across arms and mixture decomposition,
\begin{equation}
\begin{split}
\Pr(\pi_t>\pi_c \mid \Dnew,\Dhist)
&= \widetilde w_R(y_c)\;\Pr\!\big[X>Y_0\big]
\;+\; \bigl(1-\widetilde w_R(y_c)\bigr)\sum_{\ell=1}^{K}\widetilde w_\ell(y_c)\;\Pr\!\big[X>Y_\ell\big],
\end{split}
\label{eqS:pps_mix}
\end{equation}
where
\[
  X \sim \mathrm{Beta}(a_{t0}+y_t,\; b_{t0}+n_t-y_t),\quad
  Y_0 \sim \mathrm{Beta}(1+y_c,\; 1+n_c-y_c),\quad
  Y_\ell \sim \mathrm{Beta}(a_\ell+y_c,\; b_\ell+n_c-y_c).
\]
Each term $\Pr[X>Y_\bullet]$ admits the integral representation
\[
  \Pr[X>Y] \;=\; \int_0^1 \!\! \int_0^x
    \mathrm{Beta}(x\mid\alpha,\beta)\,\mathrm{Beta}(y\mid\gamma,\delta)\;dy\,dx,
\]
which can be evaluated numerically; when $\alpha$ is a positive integer, closed-form finite
series in Beta functions are available \cite{pham-gia_bayesian_1993}.

\newpage 

\section{Simulation study}
In the simulation both historical trials and prospective trial data were simulated, under various assumptions, and the operating characteristics are interpreted as unconditional measures.

Please note that for the threshold and optimal power selection rules, the retained historical pool may contain no or only one trial. In this case the robust MAP framework is not appropriate.  We therefore adapt the borrowing strategy in these cases. If no historical controls are selected, the prospective trial is analysed separately as described in Section 2 in the main manuscript. If only one historical control is selected the robust MAP prior is replaced with a \emph{robust mixture prior},
i.e. a convex combination of the vague prior \(\mathrm{Beta}(1,1)\) and the posterior \(\mathrm{Beta}(1+x_h,\,1+n_h-x_h)\) of the retained trial (see Callegaro 2023 \cite{callegaro_vaccine_2023}).

 \[
 \hat{p}_{H,\mathrm{RMP}}(\pi_{\text{new}})
 = w_R\,\mathrm{Beta}(1,1)
 + (1-w_R)\,\mathrm{Beta}\!\bigl(1+x_h,\,1+n_h-x_h\bigr),
 \qquad w_R=0.1 .
 \]

This ensures that Bayesian borrowing remains well-defined under every selection outcome while preserving comparability across selection rules.\\

The first part of this section investigated the impact of most parameters (highlighted in red) in Table 2 (from main manuscript, also shown below); Second part in monotone selection we fixed the parameters $k=8$ historical trials, with sample size per historical control arm $n_{hc}=30$, 1:1 randomization ratio and moderate between-trial heterogeneity $\tau=0.30$ to investigate different scenarios as in Table 3 (from main manuscript, also shown below). Scenario (1a +1b) can be compared to Scenario 2 to show the impact of temporal drift in historical trials, using the same prospective trial data. On the other hand Scenario (1a +1b) can also be compared to Scenario 3 to show the impact of large prospective trial, as these 2 scenarios used same historical studies pool.

% =========================
% Table 2 (Factors varied)
% =========================
\begin{table*}[!h]%
\centering%
\small
\caption*{Table 2: Factors varied in the simulation, classified as Design (D) or Assumption (A), with investigated values and description.\label{tab:scenfac1}}%
\begin{tabular*}{\linewidth}{@{\extracolsep\fill}p{2.4cm}p{1.6cm}p{3.4cm} >{\raggedright\arraybackslash}p{7.8cm}@{\extracolsep\fill}}
\toprule
\textbf{Factor} & \textbf{Type} & \textbf{Investigated values} & \textbf{Description} \\
\midrule
\textcolor{red}{$\tau$} & A & 0.10,\;0.30,\;0.50 & Between-trial heterogeneity; values motivated by Neuenschwander et al.\ (2010) \cite{neuenschwander_summarizing_2010}. \\
\textcolor{red}{$k$} & A & 4,\;8 & Number of historical trials (control arm only). \\
\textcolor{red}{$n_{hc}$} & A & 30,\;90 & Historical control sample size per trial. \\
\textcolor{red}{$n_{\text{total}}$} & D & 60,\;180 & Prospective trial total sample size. \\
\textcolor{red}{Randomization ratio} & D & 1:1,\;2:1,\;3:1 & Allocation ratio of prospective trial (treatment:control). \\
$\pi_{c}$ & A & 0.15-0.75 (step 0.05) & True prospective control response rate. \\
\bottomrule
\end{tabular*}
\begin{tablenotes}
\item \textit{Type:} D = Design; A = Assumption.
\end{tablenotes}
\end{table*}

% =========================
% Table 3 (Scenarios)
% =========================
\begin{table*}[h]%
\centering%
\small
\caption*{Table 3: Simulation scenarios for prospective trial. The scenario numbers are used for reference in the Results section.\label{tab:scenfamilies1}}%
\begin{tabular*}{\textwidth}{@{\extracolsep\fill}lp{4.2cm}p{9.8cm}@{\extracolsep\fill}}
\toprule
\textbf{Scenario} & \textbf{Name} & \textbf{Description} \\
\midrule
1a & Exchangeable & No distributional shift, $E(\pi_{hc})=E(\pi_{c})=0.20$, historical and prospective control response rates $\pi_{hc}, \pi_{c}$ are exchangeable. \\
1b &Prospective-historical distributional shift & $E(\pi_{hc})=0.20$, prospective $E(\pi_{c})$ varied from 0.15 to 0.75 (step 0.05). Represents prior-data conflict. \\
2 & Time trend & Linear logit drift ($\beta_{2}=-0.05$) across historical trials for $k=8,\;n_{hc}=30$; prospective trial fixed at $E(\pi_{c})=0.20$, earliest historical trial $E(\pi_{hc})\approx 0.14$. \\
3 & Large prospective trial & Larger prospective trial with total sample size ($n_{\text{total}}$) = 500 with smaller effects $RD=(0.0635,\,0.1152)$, calibrated so that the no-borrowing design achieves $\approx39\%$ and $84\%$ power, respectively. \\
\bottomrule
\end{tabular*}
\begin{tablenotes}
\item $RD$ denotes risk differences in prospective trial used for calibration. $\pi_{hc}$: historical control trials response rate; $\pi_{c}$: prospective trial control response rate
\end{tablenotes}
\end{table*}

\paragraph{Scenario 1a: Exchangeable}
This part supports the findings on unconditional type I error rates, power, bias (under $H_1$), square root of mean square error(RMSE, under $H_1$), effective sample size (robust MAP) reported in the main text Section 3.1, when historical control response rates and prospective control response rate are \emph{exchangeable}. The results presented are assuming $E(\pi_{hc})=E(\pi_{c})=0.20$.

\clearpage

\begin{figure}[htbp]
  \centering
  % First plot (full width)
  \includegraphics[width=\textwidth]{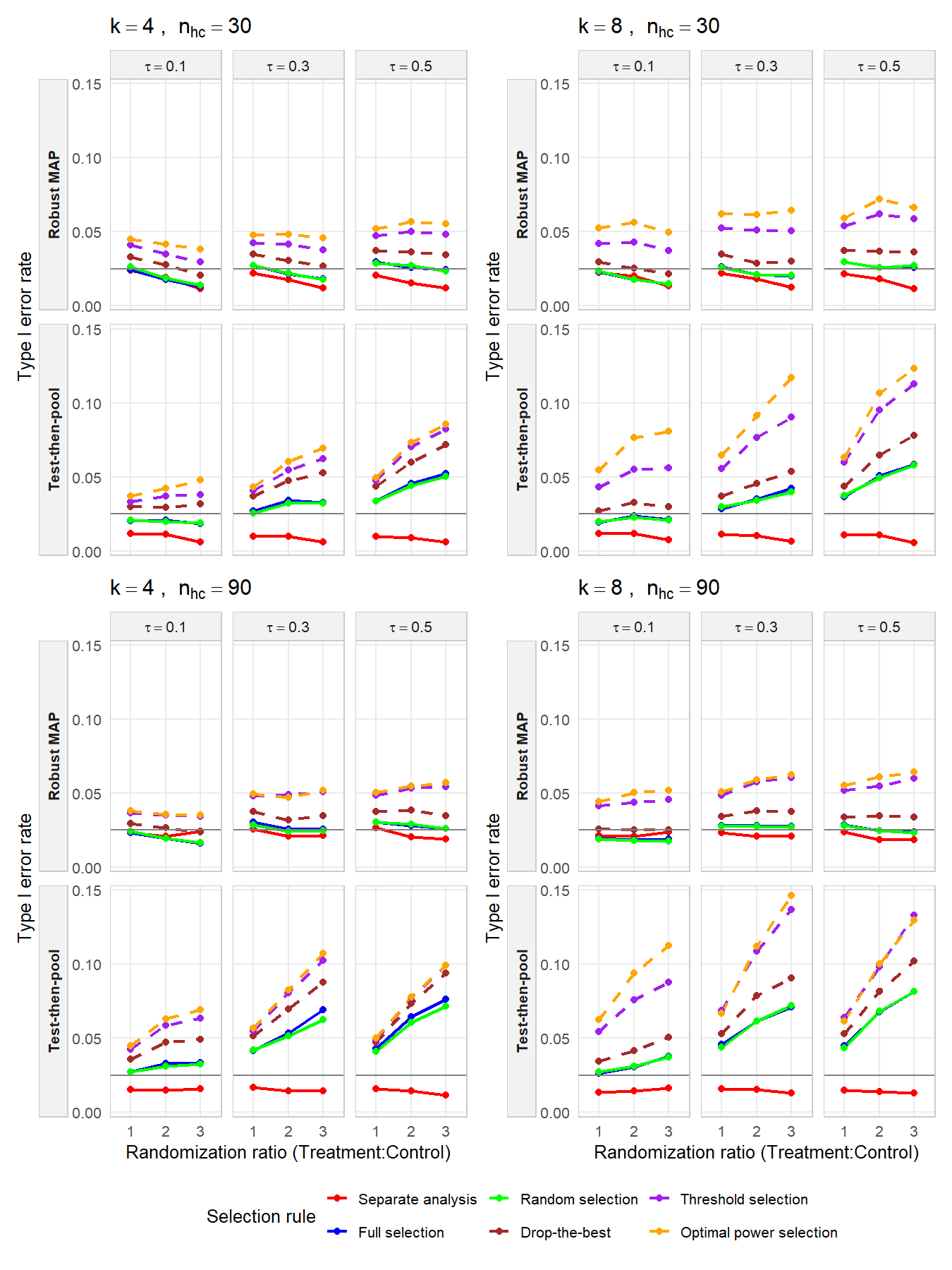}
  \caption{Type I error rates of various selection rules, to show the impact of randomization ratio (x-axis, total sample size in the prospective trial is fixed at $n_{total}=2*n_{hc}$) and between-trial heterogeneity $\tau$ (columns); Type I error rates of different numbers of historical trials $k$ and sample size per historical control arm $n_{hc}$ are provided. Dashed line: outcome-dependent selections. Results from both robust MAP and TTP are provided.}
    \label{supfig: t1e_ratio_tau1}
\end{figure}

\begin{figure}[htbp]
  \centering
  % First plot (full width)
  \includegraphics[width=\textwidth]{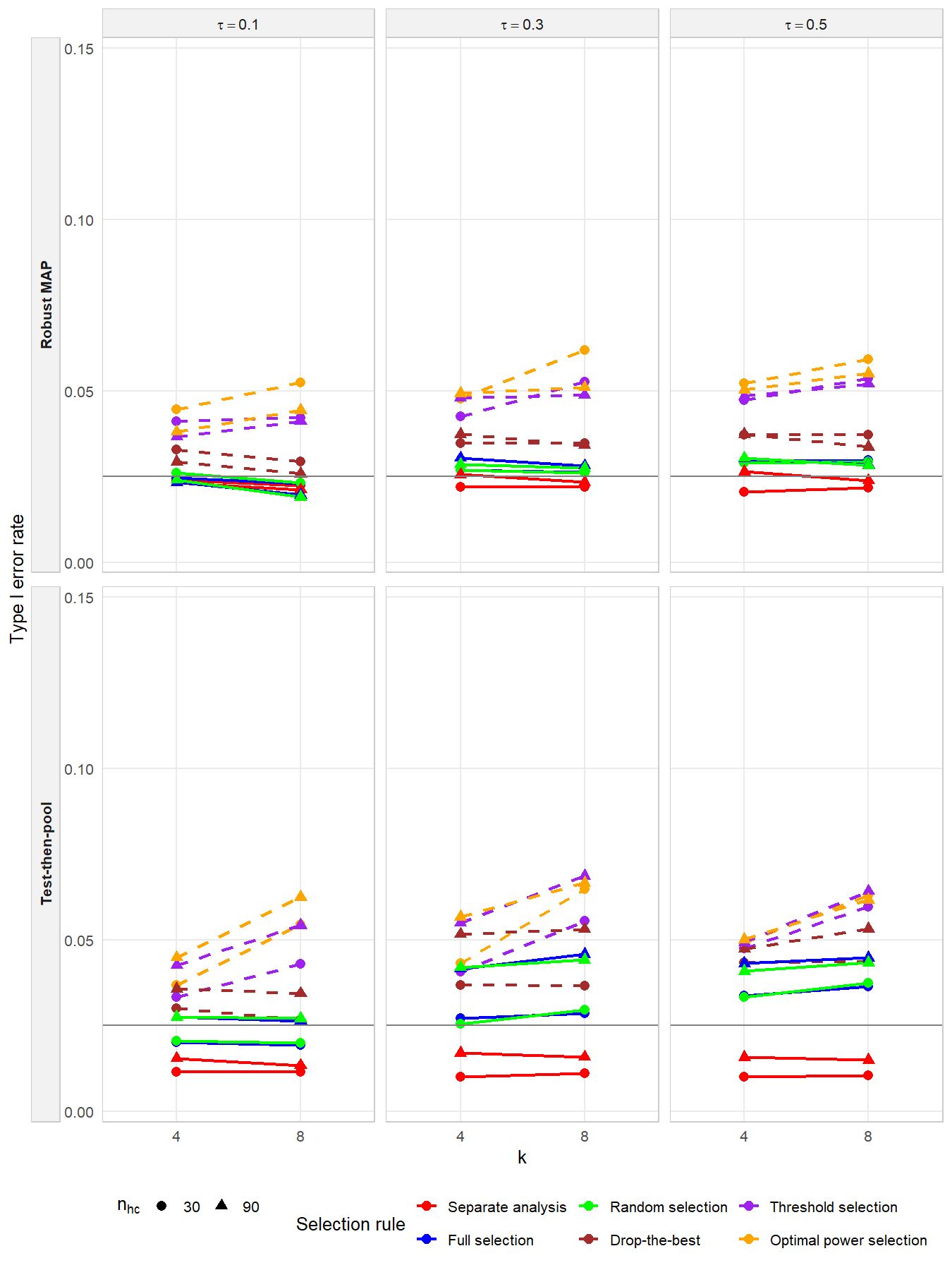}
  \caption{Type I error rates of various selection rules, to show the impact of numbers of historical trials $k$ (x-axis) and sample size per historical control arm $n_{hc}$ (circle: 30, triangle: 90); Type I error rates of different levels of between-trial heterogeneity $\tau$ are provided, with fixed 1:1 randomization ratio. Dashed line: outcome-dependent selections. Results from both robust MAP and TTP are provided. }
    \label{supfig: t1e_k_nhc1}
\end{figure}

%power
\begin{figure}[htbp]
  \centering
  % First plot (full width)
  \includegraphics[width=\textwidth]{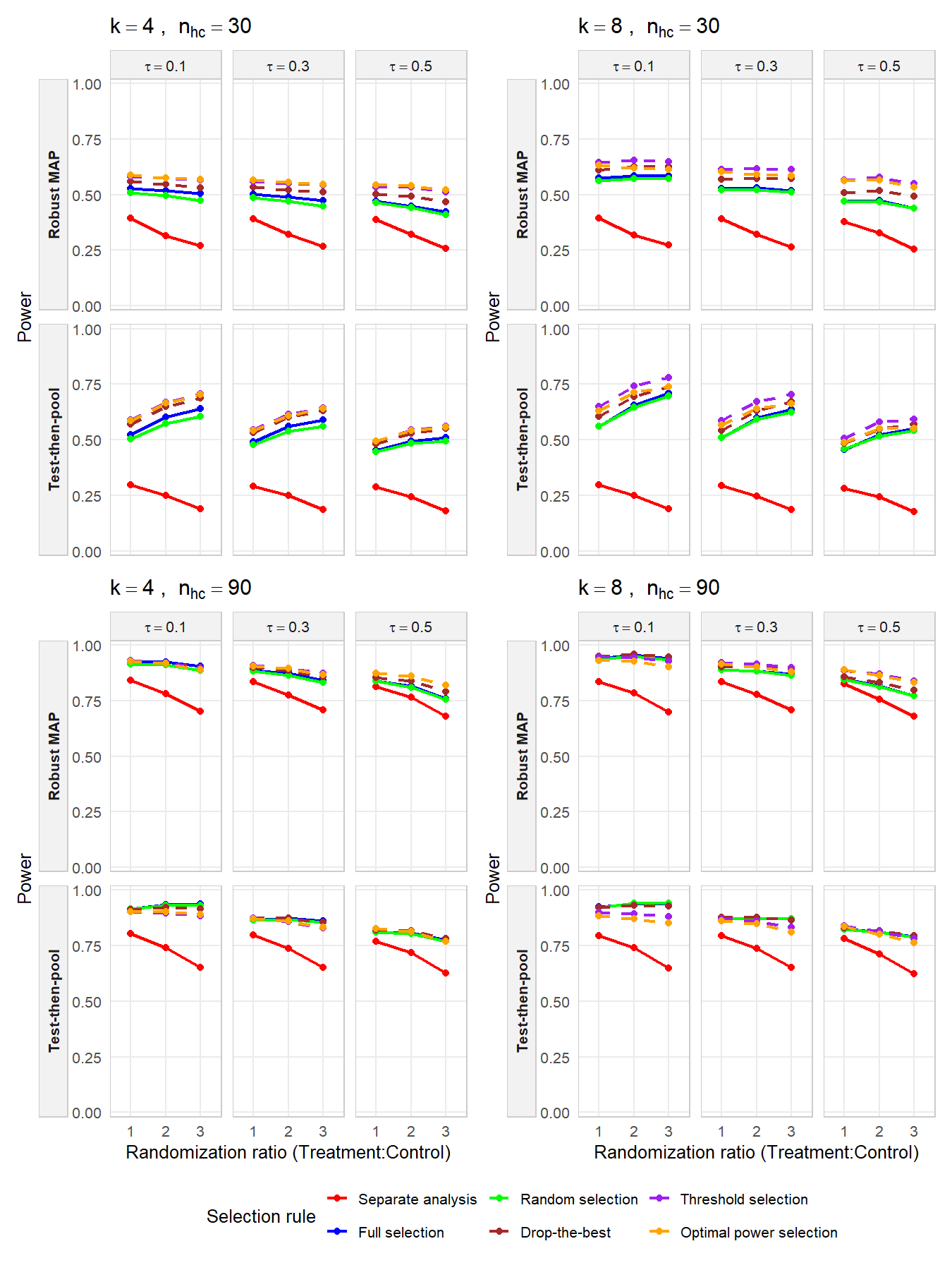}
  \caption{Power of various selection rules, to show the impact of randomization ratio (x-axis, total sample size in the prospective trial is fixed at $n_{total}=2*n_{hc}$) and between-trial heterogeneity $\tau$ (columns); Power of different numbers of historical trials $k$ and sample size per historical control arm $n_{hc}$ are provided. Dashed line: outcome-dependent selections. Results from both robust MAP and TTP are provided.}
    \label{supfig: power_ratio_tau1}
\end{figure}

\begin{figure}[htbp]
  \centering
  % First plot (full width)
  \includegraphics[width=\textwidth]{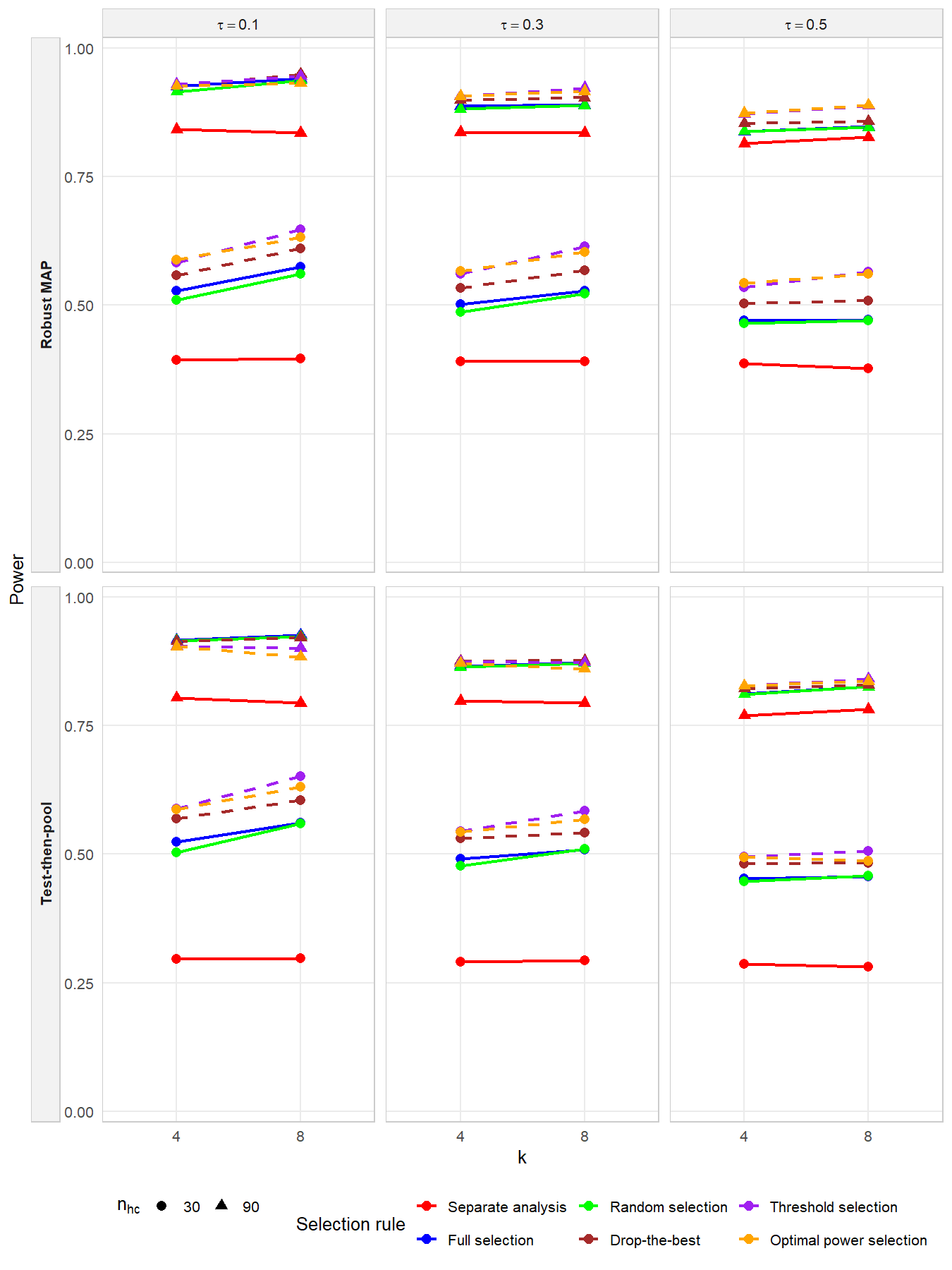}
  \caption{Power of various selection rules, to show the impact of numbers of historical trials $k$ (x-axis) and sample size per historical control arm $n_{hc}$ (circle: 30, triangle: 90); Power of different levels of between-trial heterogeneity $\tau$ are provided, with fixed 1:1 randomization ratio. Dashed line: outcome-dependent selections. Results from both robust MAP and TTP are provided.}
    \label{supfig: power_k_nhc1}
\end{figure}

%bias
\begin{figure}[htbp]
  \centering
  % First plot (full width)
  \includegraphics[width=\textwidth]{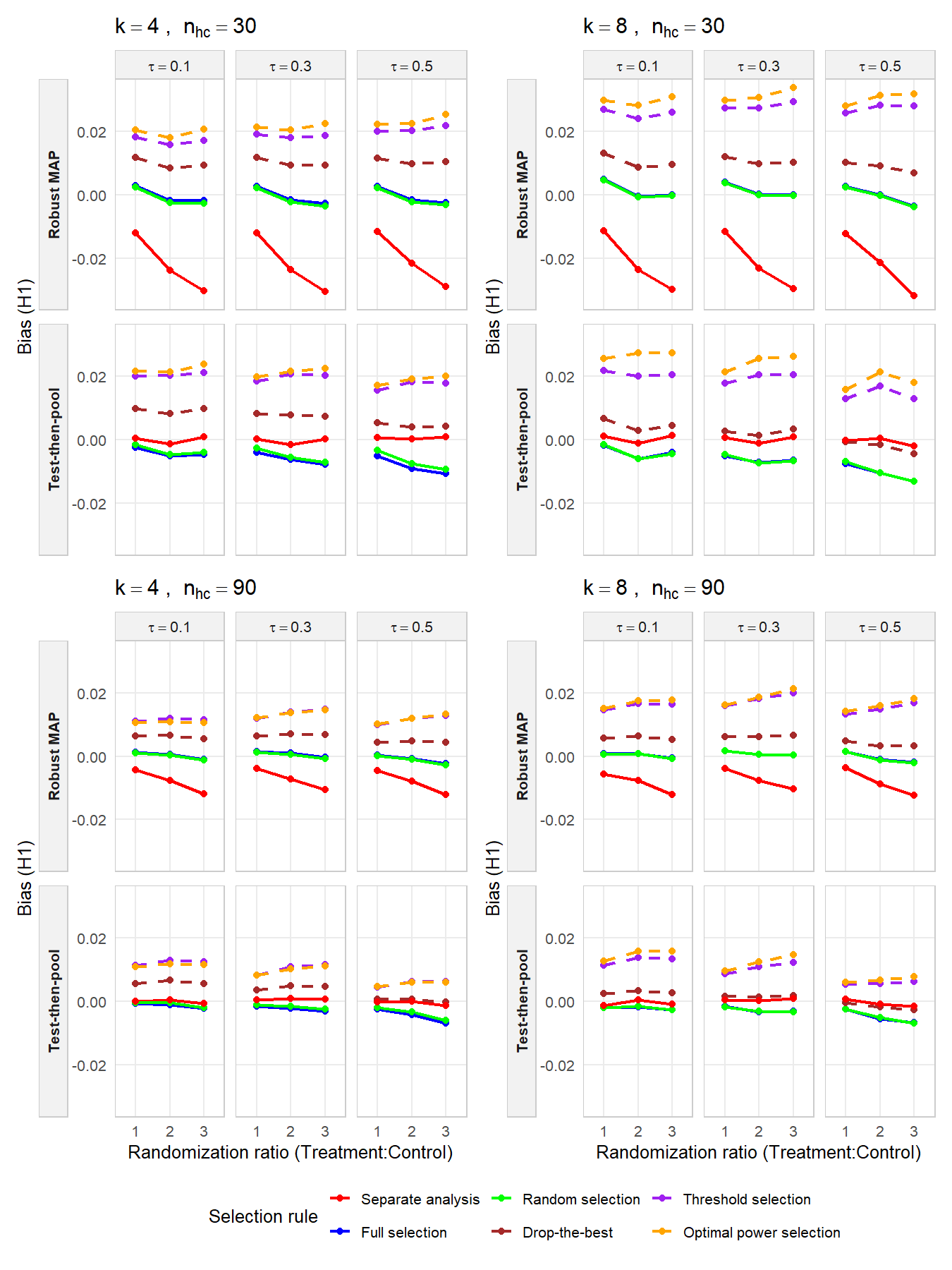}
  \caption{Bias of various selection rules under $H_1$, to show the impact of randomization ratio (x-axis, total sample size in the prospective trial is fixed at $n_{total}=2*n_{hc}$) and between-trial heterogeneity $\tau$ (columns); Bias of different numbers of historical trials $k$ and sample size per historical control arm $n_{hc}$ are provided. Dashed line: outcome-dependent selections. Results from both robust MAP and TTP are provided.}
    \label{supfig: bias_ratio_tau1}
\end{figure}

\begin{figure}[htbp]
  \centering
  % First plot (full width)
  \includegraphics[width=\textwidth]{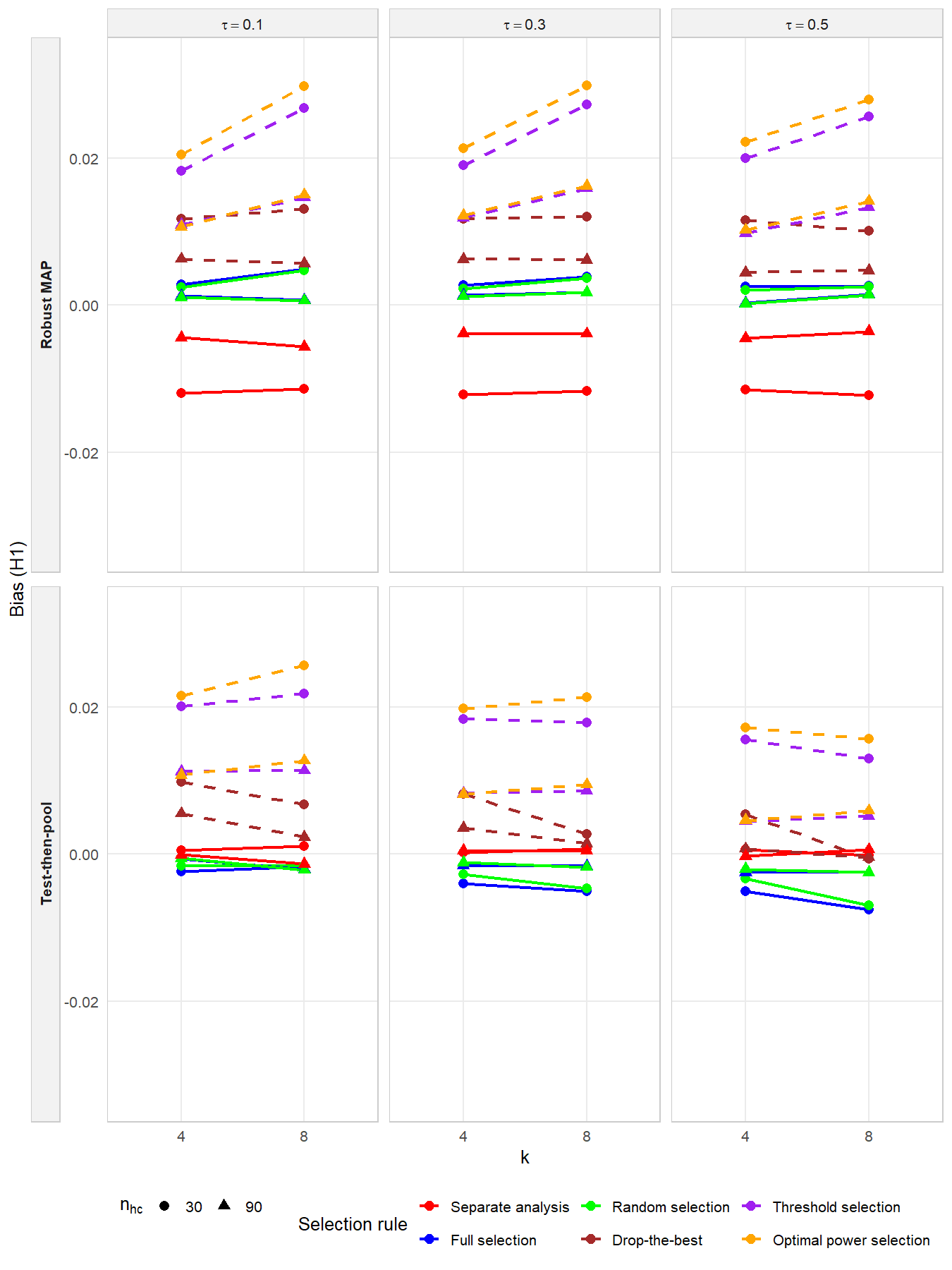}
  \caption{Bias of various selection rules under $H_1$, to show the impact of numbers of historical trials $k$ (x-axis) and sample size per historical control arm $n_{hc}$ (circle: 30, triangle: 90); Bias of different levels of between-trial heterogeneity $\tau$ are provided, with fixed 1:1 randomization ratio. Dashed line: outcome-dependent selections. Results from both robust MAP and TTP are provided.}
    \label{supfig: bias_k_nhc1}
\end{figure}
%bias, null
\begin{figure}[htbp]
  \centering
  % First plot (full width)
  \includegraphics[width=\textwidth]{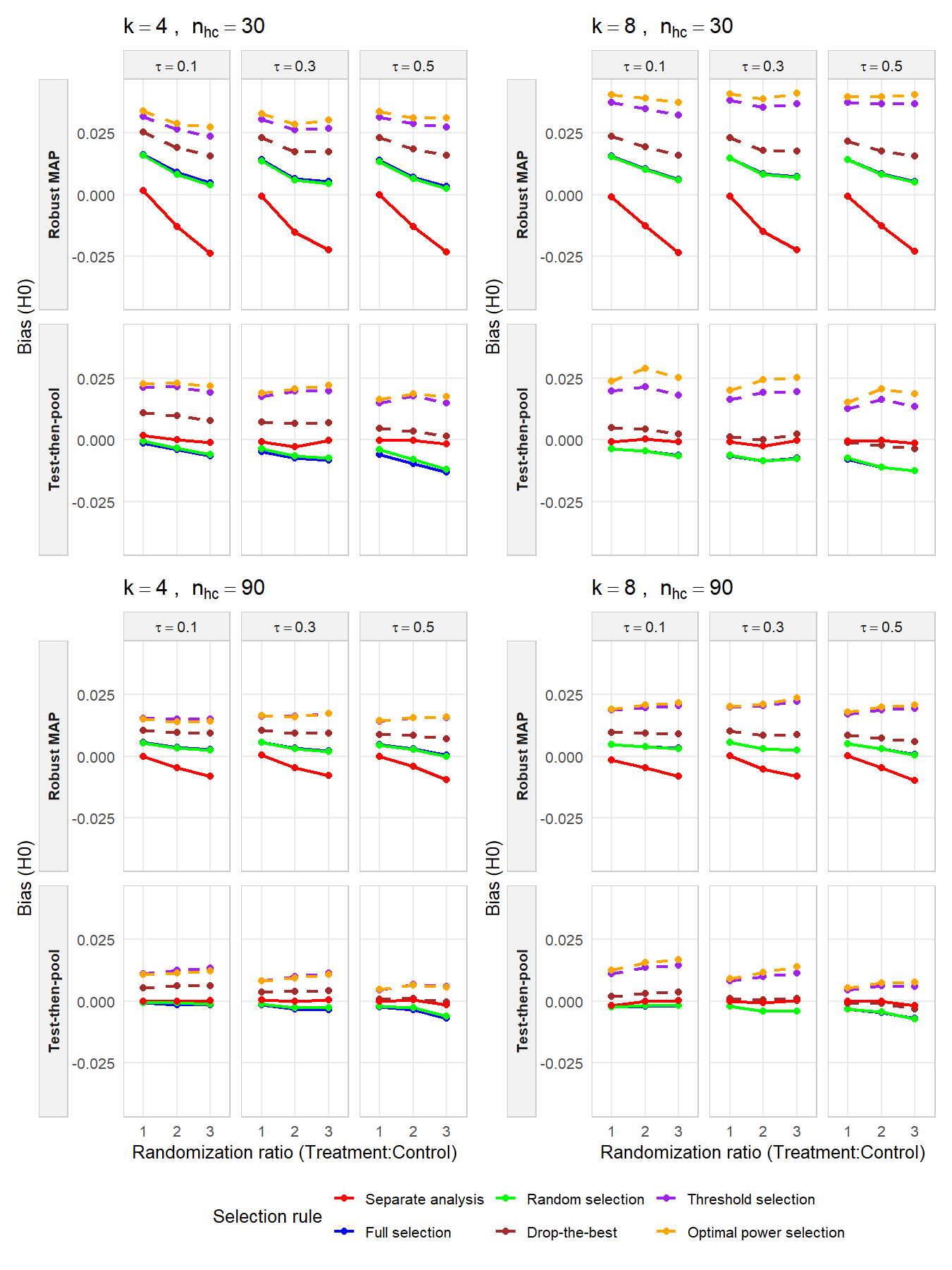}
  \caption{Bias of various selection rules under $H_0$, to show the impact of randomization ratio (x-axis, total sample size in the prospective trial is fixed at $n_{total}=2*n_{hc}$) and between-trial heterogeneity $\tau$ (columns); Bias of different numbers of historical trials $k$ and sample size per historical control arm $n_{hc}$ are provided. Dashed line: outcome-dependent selections. Results from both robust MAP and TTP are provided.}
    \label{supfig: bias_ratio_tau0}
\end{figure}

\begin{figure}[htbp]
  \centering
  % First plot (full width)
  \includegraphics[width=\textwidth]{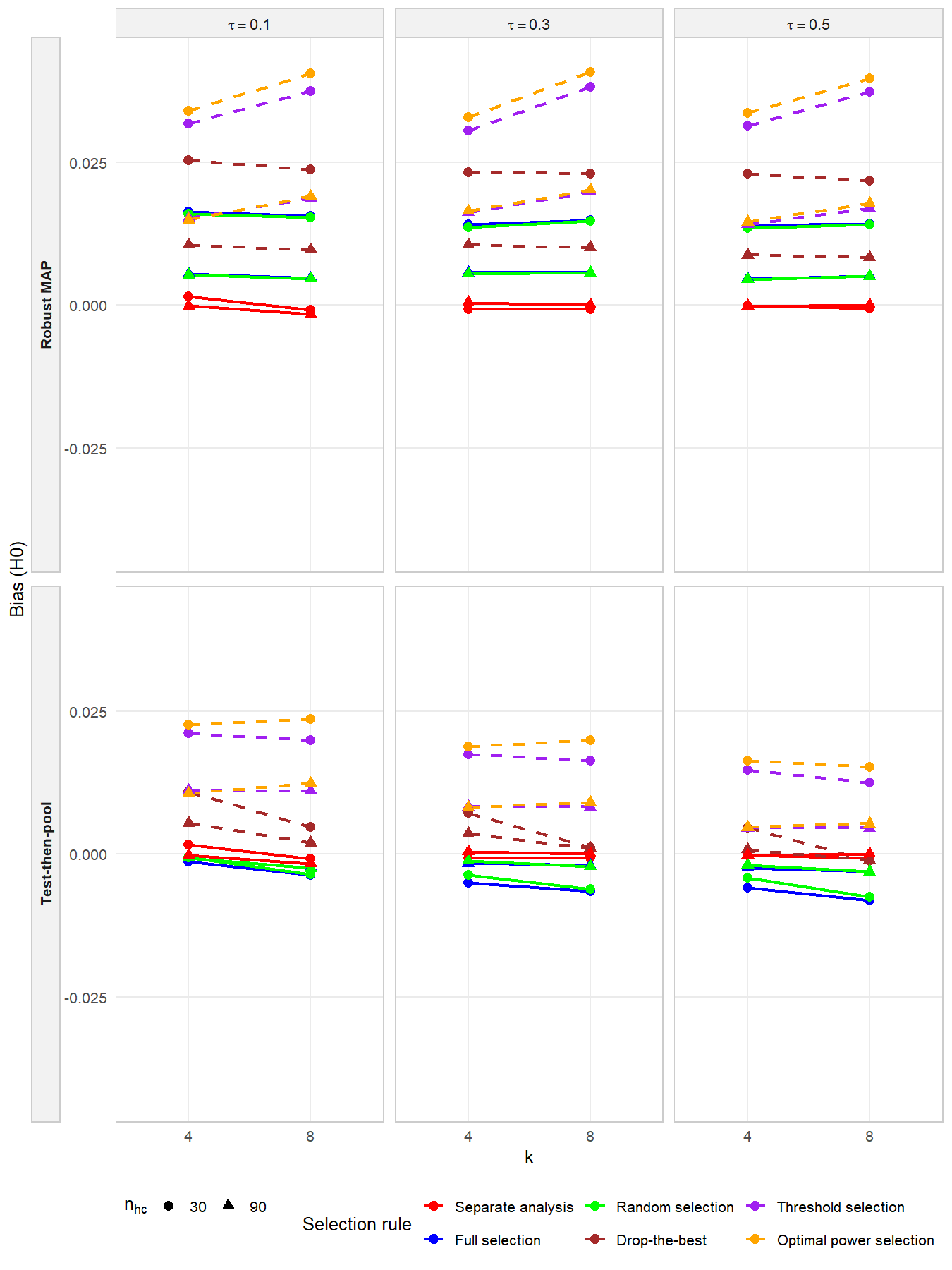}
  \caption{Bias of various selection rules under $H_0$, to show the impact of numbers of historical trials $k$ (x-axis) and sample size per historical control arm $n_{hc}$ (circle: 30, triangle: 90); Bias of different levels of between-trial heterogeneity $\tau$ are provided, with fixed 1:1 randomization ratio. Dashed line: outcome-dependent selections. Results from both robust MAP and TTP are provided.}
    \label{supfig: bias_k_nhc0}
\end{figure}
%rmse
\begin{figure}[htbp]
  \centering
  % First plot (full width)
  \includegraphics[width=\textwidth]{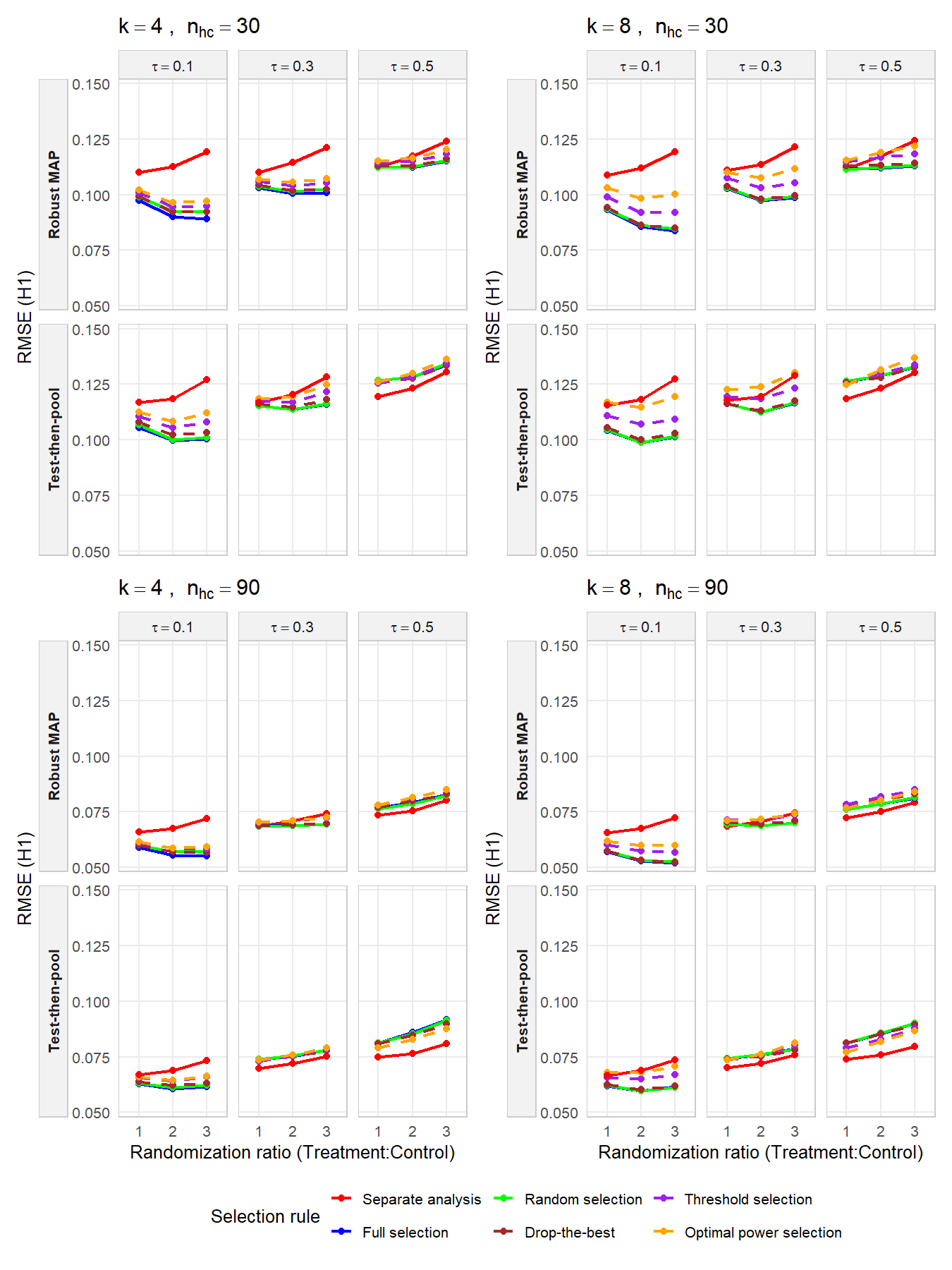}
  \caption{RMSE of various selection rules under $H_1$, to show the impact of randomization ratio (x-axis, total sample size in the prospective trial is fixed at $n_{total}=2*n_{hc}$) and between-trial heterogeneity $\tau$ (columns); RMSE of different numbers of historical trials $k$ and sample size per historical control arm $n_{hc}$ are provided. Dashed line: outcome-dependent selections. Results from both robust MAP and TTP are provided.}
    \label{supfig: rmse_ratio_tau1}
\end{figure}

\begin{figure}[htbp]
  \centering
  % First plot (full width)
  \includegraphics[width=\textwidth]{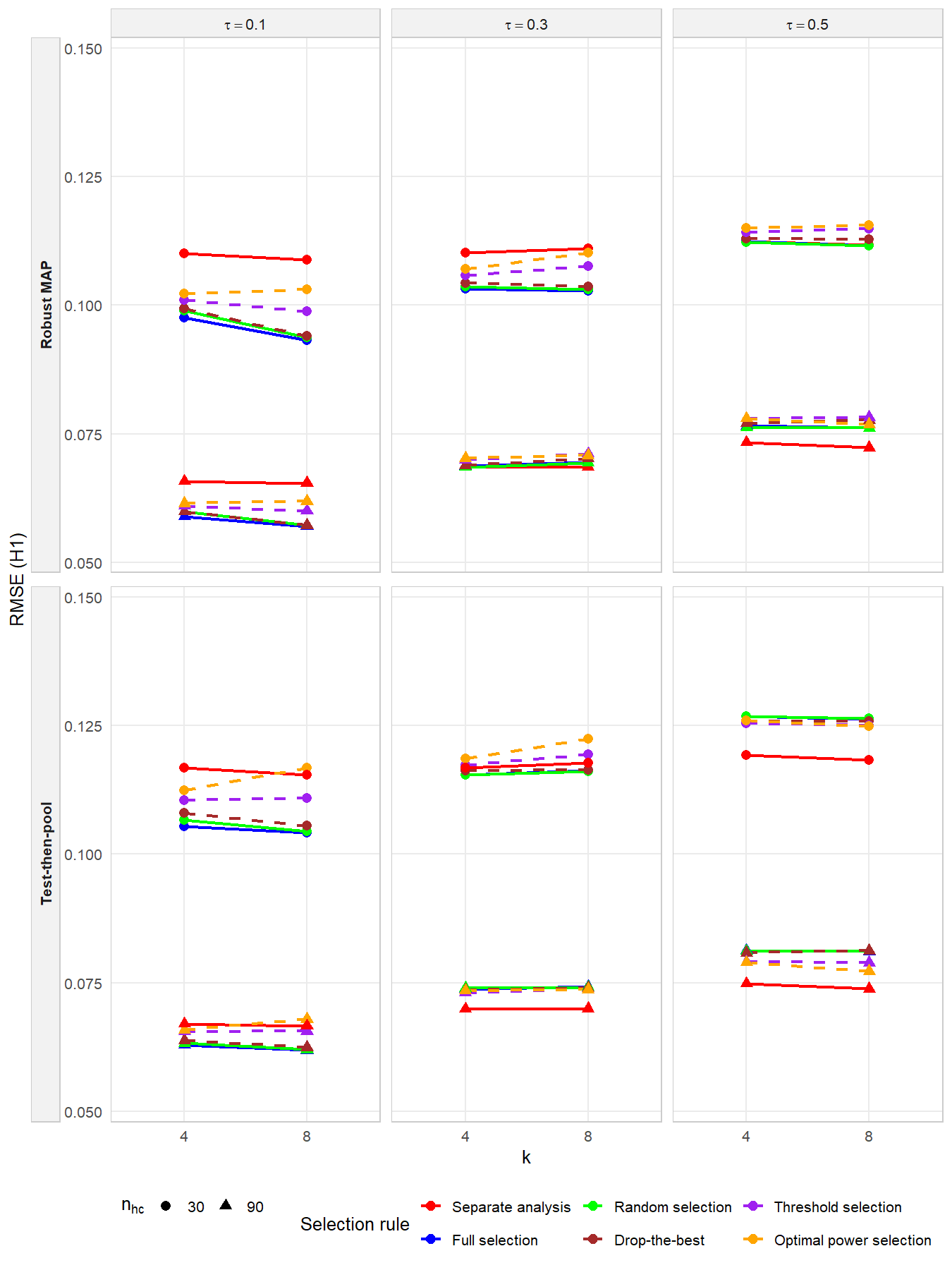}
  \caption{RMSE of various selection rules  under $H_1$, to show the impact of numbers of historical trials $k$ (x-axis) and sample size per historical control arm $n_{hc}$ (circle: 30, triangle: 90); RMSE of different levels of between-trial heterogeneity $\tau$ are provided, with fixed 1:1 randomization ratio. Dashed line: outcome-dependent selections. Results from both robust MAP and TTP are provided.}
    \label{supfig: rmse_k_nhc1}
\end{figure}

%rmse, null
\begin{figure}[htbp]
  \centering
  % First plot (full width)
  \includegraphics[width=\textwidth]{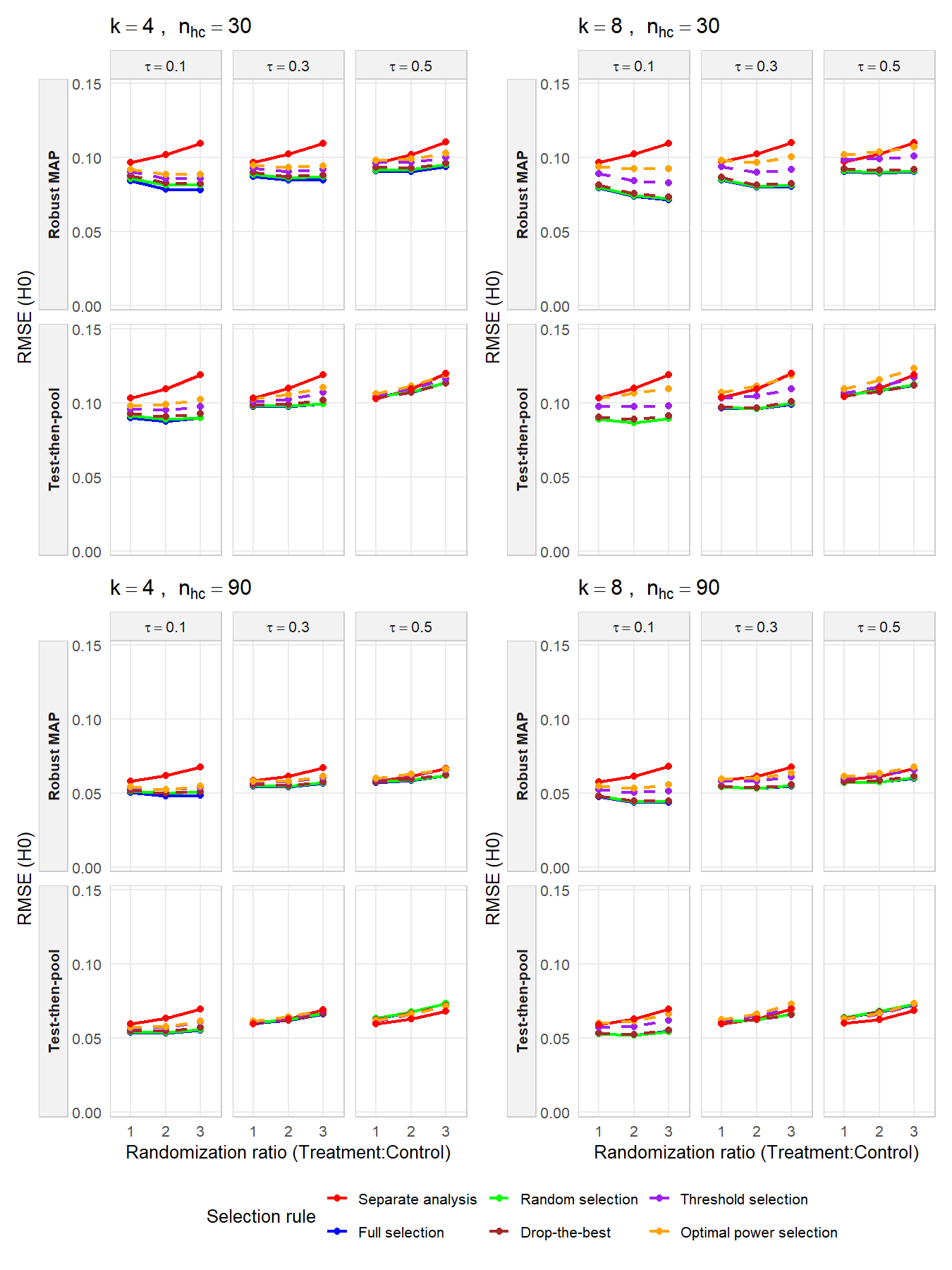}
  \caption{RMSE of various selection rules under $H_0$, to show the impact of randomization ratio (x-axis, total sample size in the prospective trial is fixed at $n_{total}=2*n_{hc}$) and between-trial heterogeneity $\tau$ (columns); RMSE of different numbers of historical trials $k$ and sample size per historical control arm $n_{hc}$ are provided. Dashed line: outcome-dependent selections. Results from both robust MAP and TTP are provided.}
    \label{supfig: rmse_ratio_tau0}
\end{figure}

\begin{figure}[htbp]
  \centering
  % First plot (full width)
  \includegraphics[width=\textwidth]{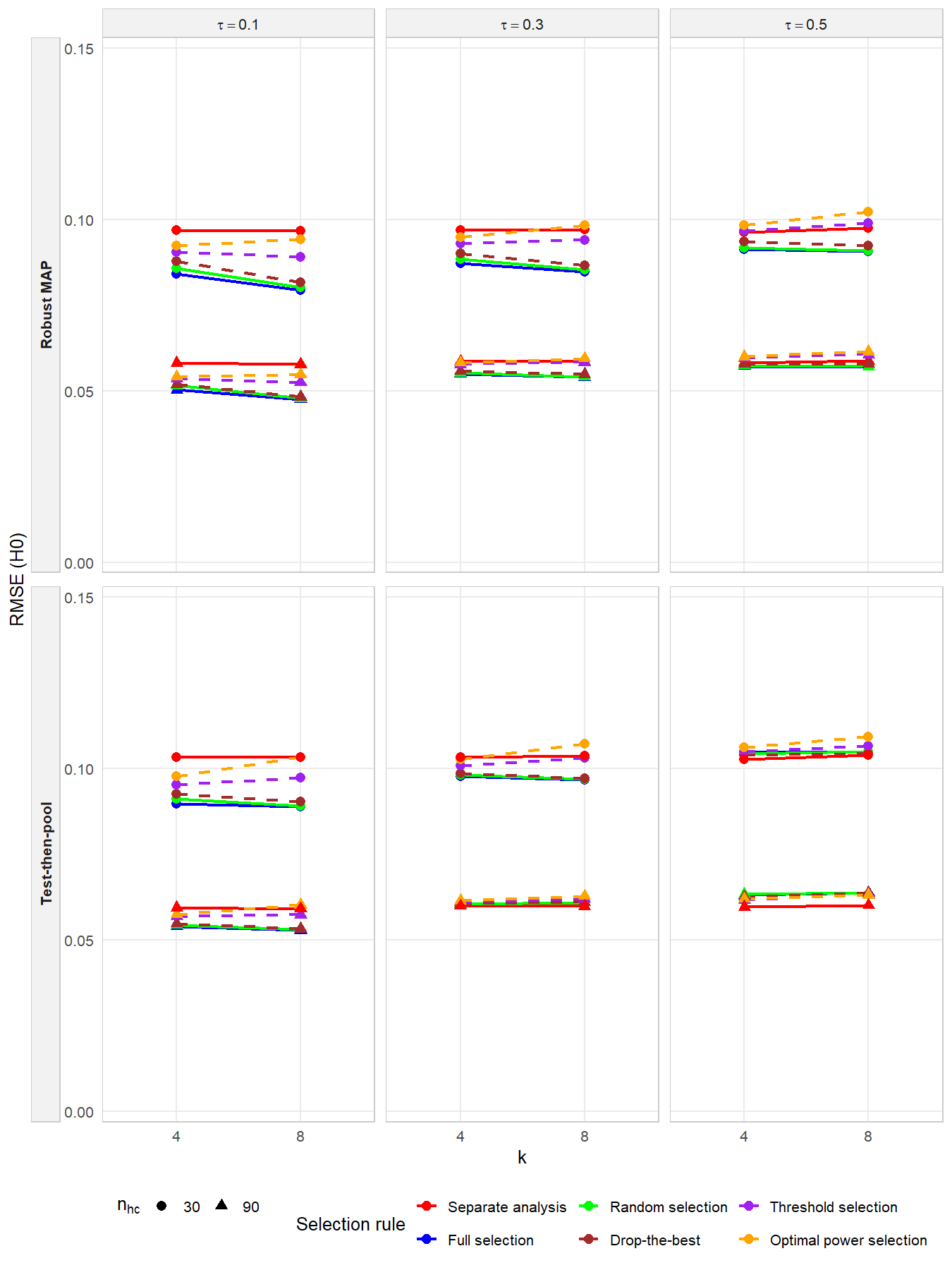}
  \caption{RMSE of various selection rules  under $H_0$, to show the impact of numbers of historical trials $k$ (x-axis) and sample size per historical control arm $n_{hc}$ (circle: 30, triangle: 90); RMSE of different levels of between-trial heterogeneity $\tau$ are provided, with fixed 1:1 randomization ratio. Dashed line: outcome-dependent selections. Results from both robust MAP and TTP are provided.}
    \label{supfig: rmse_k_nhc0}
\end{figure}

\clearpage
\begin{table}[htbp]
    \centering
    \caption{Effective Sample Size (ESS) of robust MAP prior across selection rules (excluding optimal power selection), half-normal prior for between-trial heterogeneity $\tau \;\sim\; HN(1)$ as default. $k$: numbers of historical trials; $n_{hc}$: sample size per historical control arm. Note that the total sample size in the prospective trial is fixed at $n_{total}=2*n_{hc}$.}
    \label{tab:ess_default}
    \scriptsize
    \begin{tabular}{ccccccc}
    \toprule
    $\tau$ & $k$ & $n_{hc}$ & Full selection & Random selection & Drop-the-best & Threshold selection\\
    \midrule
    0.10 & 4 & 30 & 23.52 & 16.75 & 18.76 & 19.23 \\
0.10 & 4 & 90 & 60.62 & 39.78 & 45.54 & 46.88 \\
0.10 & 8 & 30 & 51.00 & 44.45 & 52.27 & 39.85 \\
0.10 & 8 & 90 & 141.52 & 122.15 & 147.01 & 99.31 \\
0.30 & 4 & 30 & 20.79 & 15.33 & 17.85 & 18.62 \\
0.30 & 4 & 90 & 41.33 & 30.17 & 37.80 & 43.88 \\
0.30 & 8 & 30 & 40.15 & 35.68 & 44.76 & 36.19 \\
0.30 & 8 & 90 & 71.05 & 65.69 & 93.08 & 80.88 \\
0.50 & 4 & 30 & 16.81 & 13.35 & 16.42 & 18.20 \\
0.50 & 4 & 90 & 25.31 & 21.39 & 29.63 & 40.79 \\
0.50 & 8 & 30 & 26.12 & 24.29 & 34.18 & 32.41 \\
0.50 & 8 & 90 & 28.30 & 28.60 & 46.50 & 63.11 \\
    \bottomrule
    \end{tabular}
    \end{table}
\begin{table}[htbp]
    \centering
    \caption{Effective Sample Size (ESS) across selection rules (excluding optimal power selection), half-normal prior between-trial heterogeneity for $\tau \;\sim\; HN(0.5)$ as sensitivity analysis. $k$: numbers of historical trials; $n_{hc}$: sample size per historical control arm. Note that the total sample size in the prospective trial is fixed at $n_{total}=2*n_{hc}$.}
    \label{tab:ess_sensitivity}
    \scriptsize
    \begin{tabular}{ccccccc}
    \toprule
    $\tau$ & $k$ & $n_{hc}$ & Full selection & Random selection & Drop-the-best & Threshold selection\\
    \midrule
    0.10 & 4 & 30 & 32.73 & 25.66 & 28.41 & 27.19 \\
0.10 & 4 & 90 & 74.96 & 54.78 & 61.54 & 59.40 \\
0.10 & 8 & 30 & 60.03 & 53.28 & 61.61 & 50.00 \\
0.10 & 8 & 90 & 151.97 & 133.34 & 158.35 & 113.65 \\
0.30 & 4 & 30 & 29.80 & 24.00 & 27.56 & 26.57 \\
0.30 & 4 & 90 & 53.61 & 43.38 & 53.15 & 56.04 \\
0.30 & 8 & 30 & 48.92 & 44.32 & 54.44 & 46.91 \\
0.30 & 8 & 90 & 79.59 & 74.98 & 104.22 & 96.33 \\
0.50 & 4 & 30 & 25.12 & 21.48 & 26.07 & 26.26 \\
0.50 & 4 & 90 & 34.55 & 32.07 & 43.50 & 52.71 \\
0.50 & 8 & 30 & 33.65 & 31.88 & 43.77 & 43.98 \\
0.50 & 8 & 90 & 33.49 & 34.57 & 54.89 & 79.13 \\
    \bottomrule
    \end{tabular}
    \end{table}
\begin{table}[htbp]
    \centering
    \caption{Effective Sample Size (ESS) for optimal power selection by randomization ratio, half-normal prior for between-trial heterogeneity $\tau \;\sim\; HN(1)$ as default. $k$: numbers of historical trials; $n_{hc}$: sample size per historical control arm. In contrast to the other selection rules, optimal power selection depends on the design specification of the prospective trial, such as allocation ratio. Note that the total sample size in the prospective trial is fixed at $n_{total}=2*n_{hc}$.}
    \label{tab:ess_ws_default}
    \scriptsize
    \begin{tabular}{ccccc}
    \toprule
    $\tau$ & $k$ & $n_{hc}$ & Ratio & ESS \\
    \midrule
    0.10 & 4 & 30 & 1 & 19.09 \\
0.10 & 4 & 30 & 2 & 18.20 \\
0.10 & 4 & 30 & 3 & 18.35 \\
0.10 & 4 & 90 & 1 & 41.27 \\
0.10 & 4 & 90 & 2 & 42.19 \\
0.10 & 4 & 90 & 3 & 40.35 \\
0.10 & 8 & 30 & 1 & 28.10 \\
0.10 & 8 & 30 & 2 & 25.85 \\
0.10 & 8 & 30 & 3 & 24.94 \\
0.10 & 8 & 90 & 1 & 63.50 \\
0.10 & 8 & 90 & 2 & 65.14 \\
0.10 & 8 & 90 & 3 & 66.18 \\
0.30 & 4 & 30 & 1 & 19.01 \\
0.30 & 4 & 30 & 2 & 18.22 \\
0.30 & 4 & 30 & 3 & 18.52 \\
0.30 & 4 & 90 & 1 & 41.91 \\
0.30 & 4 & 90 & 2 & 40.69 \\
0.30 & 4 & 90 & 3 & 39.56 \\
0.30 & 8 & 30 & 1 & 26.68 \\
0.30 & 8 & 30 & 2 & 24.75 \\
0.30 & 8 & 30 & 3 & 23.99 \\
0.30 & 8 & 90 & 1 & 55.03 \\
0.30 & 8 & 90 & 2 & 55.05 \\
0.30 & 8 & 90 & 3 & 54.98 \\
0.50 & 4 & 30 & 1 & 19.11 \\
0.50 & 4 & 30 & 2 & 18.55 \\
0.50 & 4 & 30 & 3 & 18.80 \\
0.50 & 4 & 90 & 1 & 42.36 \\
0.50 & 4 & 90 & 2 & 40.82 \\
0.50 & 4 & 90 & 3 & 40.38 \\
0.50 & 8 & 30 & 1 & 25.53 \\
0.50 & 8 & 30 & 2 & 24.12 \\
0.50 & 8 & 30 & 3 & 23.46 \\
0.50 & 8 & 90 & 1 & 49.13 \\
0.50 & 8 & 90 & 2 & 48.51 \\
0.50 & 8 & 90 & 3 & 47.92 \\
    \bottomrule
    \end{tabular}
    \end{table}
\begin{table}[htbp]
    \centering
    \caption{Effective Sample Size (ESS) for optimal power selection by randomization ratio, under the alternative hypothesis; half-normal prior for between-trial heterogeneity $\tau \;\sim\; HN(0.5)$ as sensitivity analysis. $k$: numbers of historical trials; $n_{hc}$: sample size per historical control arm. In contrast to the other selection rules, optimal power selection depends on the design specification of the prospective trial, such as allocation ratio. Note that the total sample size in the prospective trial is fixed at $n_{total}=2*n_{hc}$.}
    \label{tab:ess_ws_sensitivity}
    \scriptsize
    \begin{tabular}{ccccc}
    \toprule
    $\tau$ & $k$ & $n_{hc}$ & Ratio & ESS \\
    \midrule
    0.10 & 4 & 30 & 1 & 26.54 \\
0.10 & 4 & 30 & 2 & 25.88 \\
0.10 & 4 & 30 & 3 & 25.72 \\
0.10 & 4 & 90 & 1 & 55.37 \\
0.10 & 4 & 90 & 2 & 56.91 \\
0.10 & 4 & 90 & 3 & 55.19 \\
0.10 & 8 & 30 & 1 & 38.13 \\
0.10 & 8 & 30 & 2 & 35.75 \\
0.10 & 8 & 30 & 3 & 34.88 \\
0.10 & 8 & 90 & 1 & 79.57 \\
0.10 & 8 & 90 & 2 & 81.29 \\
0.10 & 8 & 90 & 3 & 82.12 \\
0.30 & 4 & 30 & 1 & 26.32 \\
0.30 & 4 & 30 & 2 & 25.75 \\
0.30 & 4 & 30 & 3 & 25.75 \\
0.30 & 4 & 90 & 1 & 54.97 \\
0.30 & 4 & 90 & 2 & 54.54 \\
0.30 & 4 & 90 & 3 & 53.47 \\
0.30 & 8 & 30 & 1 & 37.06 \\
0.30 & 8 & 30 & 2 & 35.00 \\
0.30 & 8 & 30 & 3 & 34.27 \\
0.30 & 8 & 90 & 1 & 71.81 \\
0.30 & 8 & 90 & 2 & 72.04 \\
0.30 & 8 & 90 & 3 & 71.83 \\
0.50 & 4 & 30 & 1 & 26.41 \\
0.50 & 4 & 30 & 2 & 26.00 \\
0.50 & 4 & 30 & 3 & 26.04 \\
0.50 & 4 & 90 & 1 & 54.80 \\
0.50 & 4 & 90 & 2 & 53.89 \\
0.50 & 4 & 90 & 3 & 53.50 \\
0.50 & 8 & 30 & 1 & 36.65 \\
0.50 & 8 & 30 & 2 & 35.08 \\
0.50 & 8 & 30 & 3 & 34.45 \\
0.50 & 8 & 90 & 1 & 66.89 \\
0.50 & 8 & 90 & 2 & 66.46 \\
0.50 & 8 & 90 & 3 & 65.79 \\
    \bottomrule
    \end{tabular}
    \end{table}

\clearpage

\paragraph{Monotone selection}
Monotone selection is a stricter selection rule introduced in Section 5 of the main text as a possible way to mitigate outcome-dependent selection bias. The performance measures of monotone selection implemented on optimal power selected trial are shown in Supplementary Table S5 and Supplementary Figure S9-S10. For the ease of interpretation only results of $k=8$ historical trials, with sample size per historical control arm $n_{hc}=30$, and moderate between-trial heterogeneity $\tau=0.30$ are provided.

\begin{table}[htbp]
\centering
\small
  \setlength\tabcolsep{2pt}
    \caption{As Table 5 ($k=8$ historical trials, with sample size per historical control arm $n_{hc}=30$, and moderate between-trial heterogeneity $\tau=0.30$), with an additional row reporting the monotone selection constraint applied to the optimal power selection. \emph{Monotone selection} (red): if a later trial is excluded, all earlier trials are also excluded.}
    \label{tab:mono}

\begin{tabular}{lcccccccccccc}
  \toprule
  & \multicolumn{4}{c}{\textbf{Exchangeable}} 
  & \multicolumn{4}{c}{\textbf{Time-trend}} 
  & \multicolumn{4}{c}{\textbf{Large prospective trial}}\\
  \cmidrule(lr){2-5}\cmidrule(lr){6-9}\cmidrule(lr){10-13}
  \multirow{-2}{*}{\textbf{Selection rules}} 
    & T1E & Power & Bias & RMSE 
    & T1E & Power & Bias & RMSE 
    & T1E & Power & Bias & RMSE\\
  \midrule
    Separate analysis           & 2.20 & 39.05 & $-1.17$ &11.09  & 2.20 & 39.05 & $-1.17$ &11.09  & 2.57 & 38.87 & $-0.04$ &3.85 \\
    Full selection               & 2.63 & 52.75 &  0.38   &10.28  & 3.95 & 58.32 &  1.85   &10.52  & 2.83 & 42.91 &  0.18   &3.90 \\
    Random selection             & 2.60 & 52.21 &  0.37   &10.31  & 3.95 & 57.70 &  1.78   &10.52  & 2.79 & 42.62 &  0.17   &3.89 \\
    Drop-the-best                & 3.48 & 56.80 &  1.21   &10.36  & 5.27 & 61.92 &  2.62   &10.73  & 3.28 & 45.98 &  0.41   &3.92 \\
    Threshold selection          & 5.26 & 61.43 &  2.73   &10.76  & 6.54 & 63.96 &  3.38   &11.01  & 4.29 & 49.38 &  0.80   &3.95 \\
    Optimal power selection      & 6.18 & 60.33 &  2.98   &11.02  & 7.33 & 60.67 &  3.47   &11.34  & 4.35 & 49.03 &  0.79   &3.95 \\
  \textcolor{red}{Monotone selection} &3.63  &48.02  &0.43    &10.95  &3.62  &45.97  &0.15    &11.06  &3.09  &42.95  &0.30    &3.88 \\
  \bottomrule
\end{tabular}%
\begin{tablenotes}
\item[$^{\rm a}$] Exchangeable: true risk difference $RD=0.20$.
\item[$^{\rm b}$] Time-trend: linear drift in the historical data.
\item[$^{\rm c}$] Large prospective trial: prospective trial with $RD=0.0635$ with $n=250$ per arm.
\item $RD$: treatment effect measure, risk differences between the treatment and control arm.\\ All values are \textbf{percentages}. T1E: type I error rate at nominal one-sided level $2.5\%$. Bias: percentage-point deviation of posterior mean RD from the truth, under $H_1$, calculated by $\displaystyle \frac{1}{n_{\text{sim}}}\sum_{r=1}^{n_{\text{sim}}} \bigl(\hat{RD}^{(r)}-RD\bigr)\times 100$. RMSE: root mean square error, under $H_1$, calculated by $\displaystyle \sqrt{\frac{1}{n_{\text{sim}}}\sum_{r=1}^{n_{\text{sim}}}
              \bigl(\hat{RD}^{(r)}-RD\bigr)^2}\times 100$..
\end{tablenotes}
\end{table}

% --- Null hypothesis results ---
\begin{figure}[htbp]
  \centering
  \includegraphics[width=0.8\textwidth]{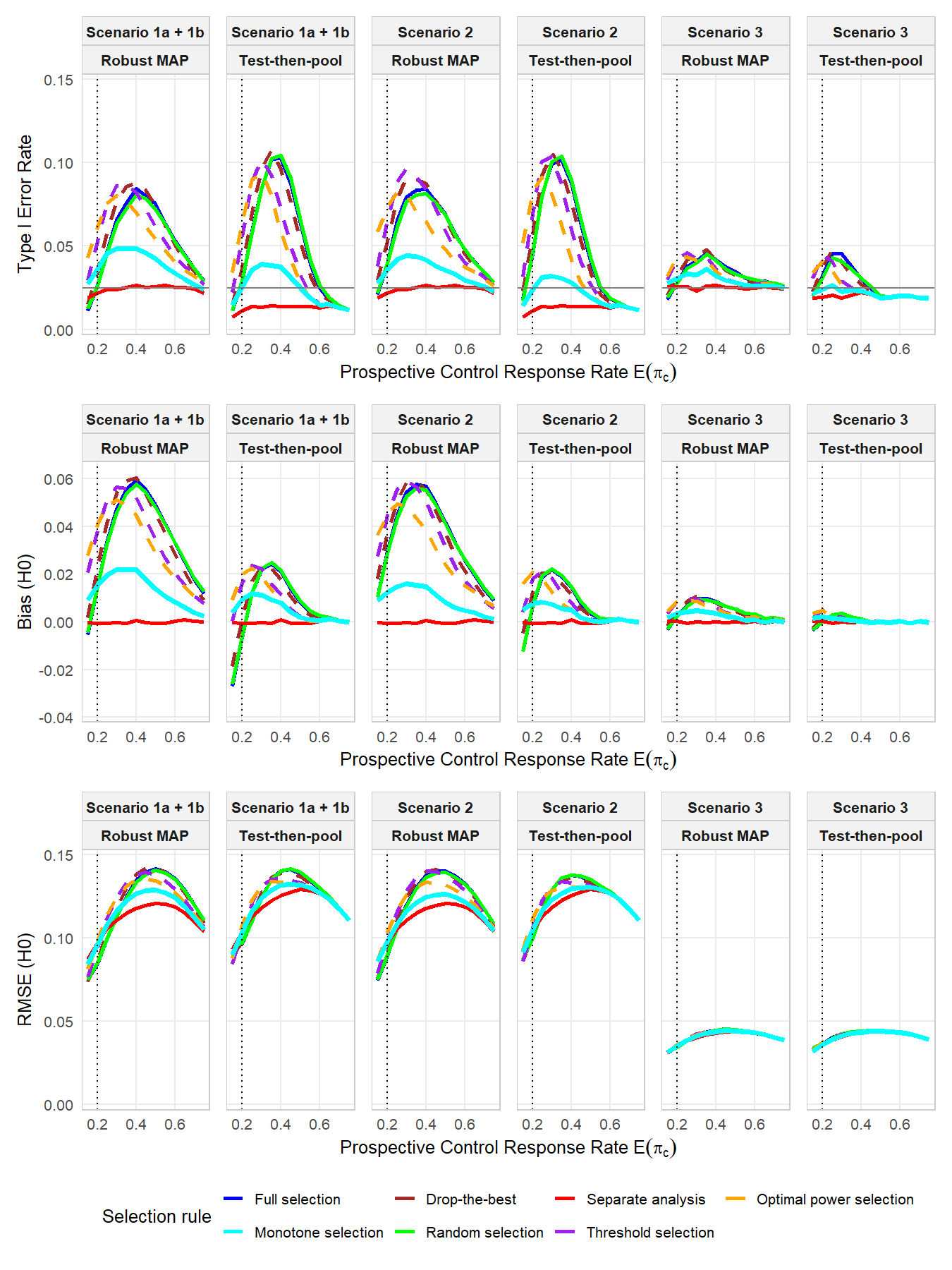}
  \caption{Performance under the null hypothesis with $k=8$ historical trials, with sample size per historical control arm $n_{hc}=30$, 
  and moderate heterogeneity between historical trials ($\tau=0.30$). Scenario 1a + 1b : true risk difference $RD=0.20$. Scenario 2 Time-trend: linear drift in historical data. Scenario 3 Large prospective trial: $RD=0.0635$ with $n=250$ per arm. 
  Panels show Type~I error rate, Bias (H0), and RMSE (H0). 
  x-axis: Prospective control response rate; the vertical line at $x=0.20$ indicates exchangeability (Scenario 1a). 
  Dashed line: outcome-dependent selections. 
  Cyan solid line: Monotone selection, mitigating the impact of optimal power selection (orange dashed line).}
  \label{supfig:OC_null_mono}
\end{figure}

% --- Alternative hypothesis results ---
\begin{figure}[htbp]
  \centering
  \includegraphics[width=0.8\textwidth]{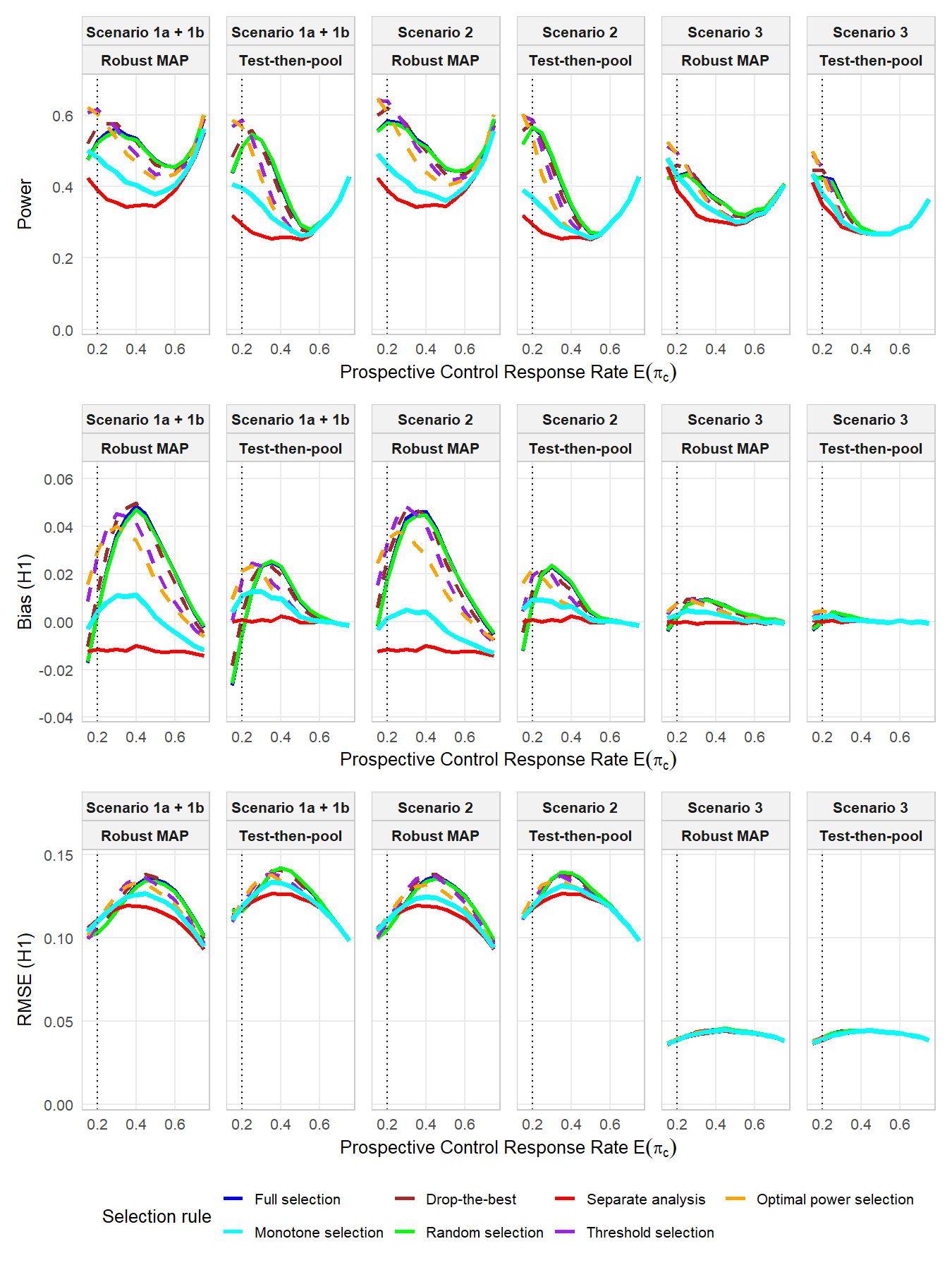}
  \caption{Performance under the alternative hypothesis with $k=8$ historical trials, with sample size per historical control arm $n_{hc}=30$, 
  and moderate heterogeneity between historical trials ($\tau=0.30$). Scenario 1a + 1b : true risk difference $RD=0.20$. Scenario 2 Time-trend: linear drift in historical data. Scenario 3 Large prospective trial: $RD=0.0635$ with $n=250$ per arm.
  Panels show Power, Bias (H1), and RMSE (H1). 
  x-axis: Prospective control response rate; the vertical line at $x=0.20$ indicates exchangeability (Scenario 1a).
  Dashed line: outcome-dependent selections. 
  Cyan solid line: Monotone selection, mitigating the impact of optimal power selection (orange dashed line).}
  \label{supfig:OC_alt_mono}
\end{figure}

\newpage
\section{Additional information for case study}
In this section we present additional information for the case study in the ankylosing spondylitis (AS) trial data originally from Baeten et al.\cite{baeten_anti-interleukin-17a_2013}. In the main manuscript the case study is described in Section 4 more in detail. Supplementary Figure S11 shows which of the 8 historical studies have been selected depending on the selection rule. Supplementary Table S6 shows the results when applying a frequentist analysis TTP to the selected trials (see methods in main text Section 2.1). The two-sided pre-test at significance level 0.1 was implemented using Fisher’s exact test. In the second step a one-sided Fisher test at $\alpha = 0.025$ was used to compare the treatment and control. The estimates and 95\% confidence intervals of risk differences are obtained from two-sample test for equality of proportions with Yates continuity correction (large-sample chi-square test, \texttt{prop.test()} function), as the Fisher's Exact test only provides odds ratio. In Supplementary Figure S12 additional conditional results based on the same historical studies with varying prospective settings are presented.

\begin{figure}[htbp]
  \centering
  \includegraphics[width=\textwidth]{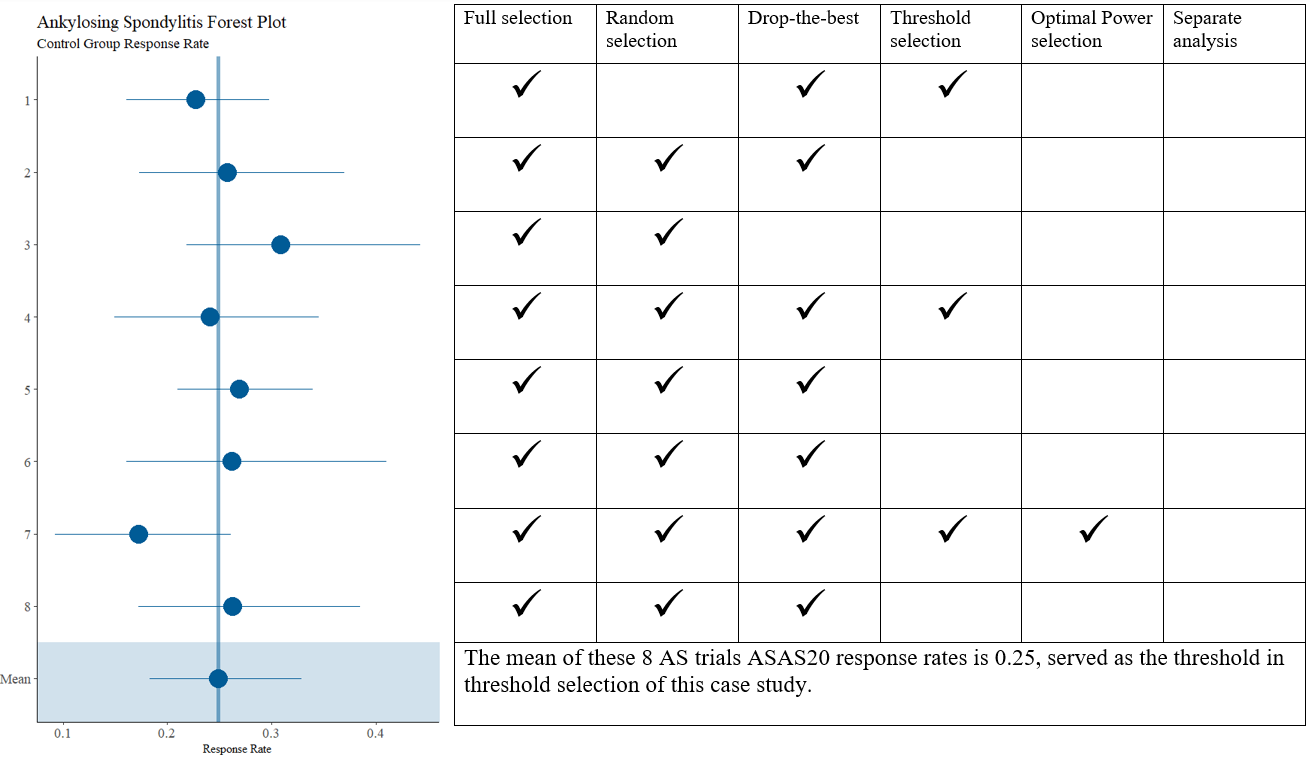}
  \caption{Selection results in the case study. Left: Forest plot of AS data, centered at 0.25; Right: selected trials with indicators ($\checkmark$)}
    \label{supfig: casestudyselection}
\end{figure}

\begin{figure}[htbp]
  \centering

  \begin{subfigure}{\textwidth}
    \includegraphics[width=\textwidth]{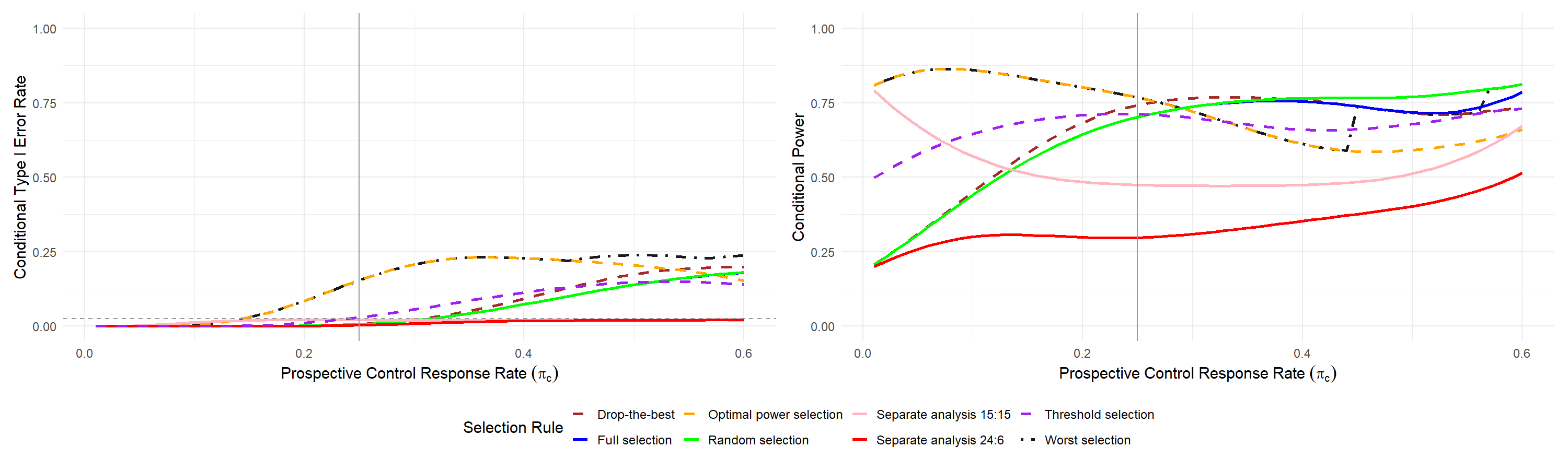}
    \caption{Conditional type I error and conditional power for $n_\text{total}=30$.}
    \label{fig:oc-case-30}
  \end{subfigure}\par\medskip

  \begin{subfigure}{\textwidth}
    \includegraphics[width=\textwidth]{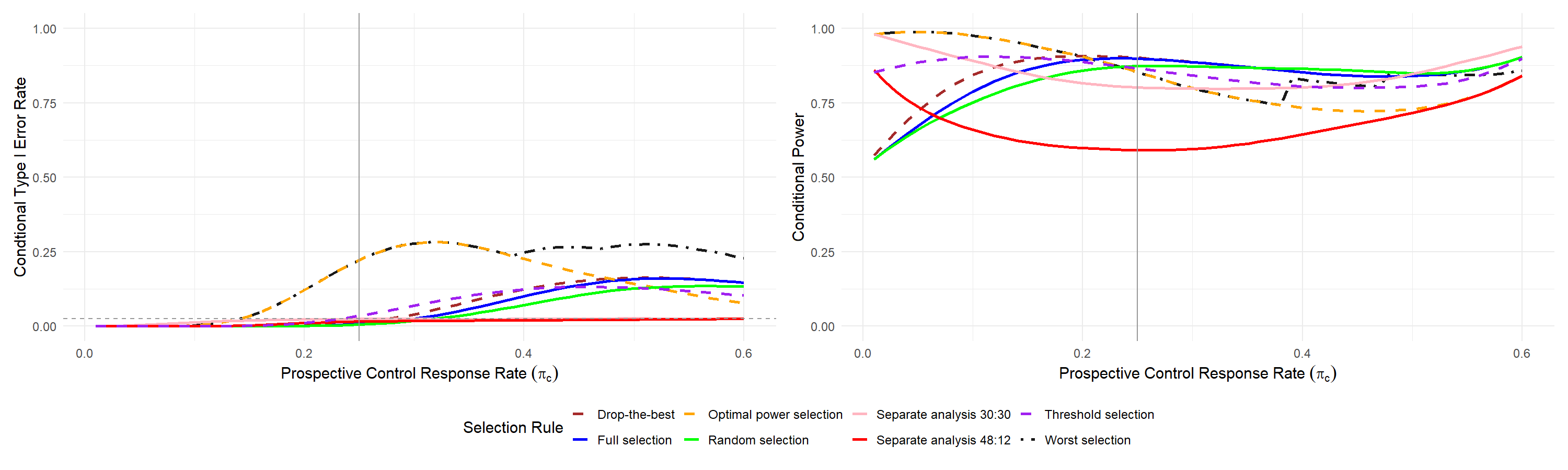}
    \caption{Conditional type I error and conditional power for $n_\text{total}=60$.}
    \label{fig:oc-case-60}
  \end{subfigure}\par\medskip

  \begin{subfigure}{\textwidth}
    \includegraphics[width=\textwidth]{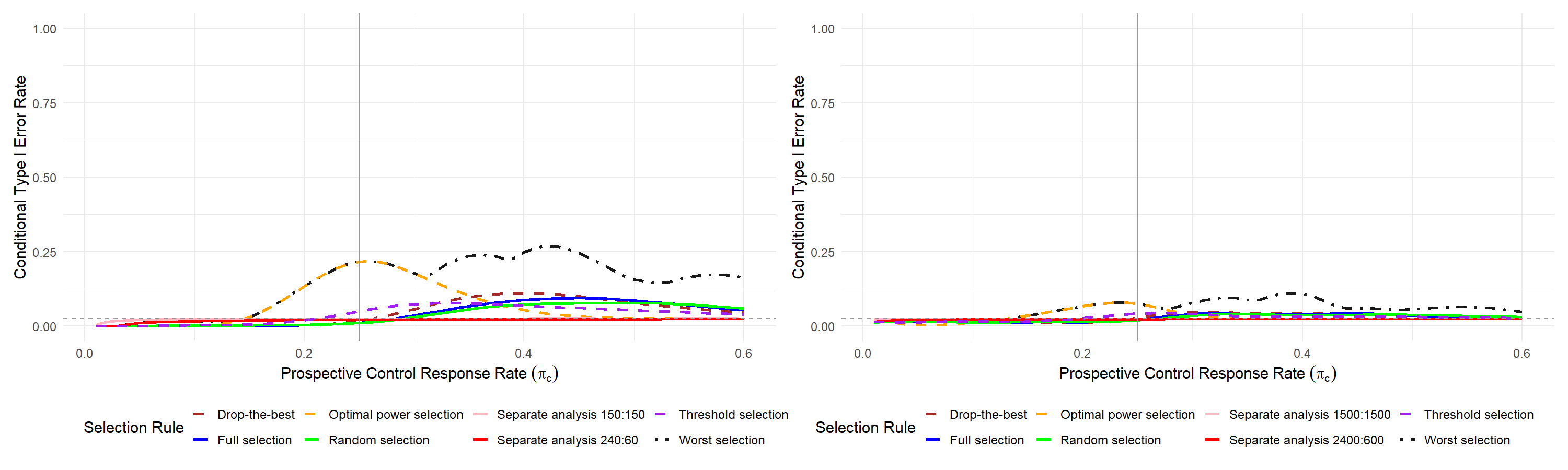}
    \caption{Conditional type I error for $n_\text{total}=300,3000$.}
    \label{fig:oc-case-300-3000}
  \end{subfigure}

  \caption{Conditional operating characteristics of different selection rules based on the same historical pool and varying prospective settings. (a) Conditional type I error rates and power when prospective trial planned to have total sample size $n_\text{total}=30$. (b) Conditional type I error rates and power when prospective trial planned to have total sample size $n_\text{total}=60$. (c) Conditional type I error rates when prospective trial planned to have total sample size $n_\text{total}=300$ (left) and $n_\text{total}=3000$ (right), power is not shown here as under this sample size the power will be very close to 1. The black dashed line marks the worst (max conditional type I error) selection.}
  \label{supfig: OC case study comparison}
\end{figure}

\begin{center}
\begin{table*}[!ht]
\caption{Results of applying various selection rules to the ankylosing spondylitis (AS) case study trial data, analyzed with TTP.\label{tab:case_study_freq}}
\begin{tabular*}{\textwidth}{@{\extracolsep\fill} l c c c @{}}
\toprule
\textbf{Selection rule} & \textbf{Estimate} & \textbf{95\% CI} & \textbf{$p$-value} \\
\midrule
Full selection                & 0.337 & (0.114, 0.559) & $<0.001$ \\ %6.5e\text{-}04
Random selection                & 0.328 & (0.105, 0.552) & $<0.001$  \\ %1.1e\text{-}03
Drop-the-best                & 0.350 & (0.128, 0.573) & $<0.001$ \\ %3.7e\text{-}04
Threshold selection                & 0.401 & (0.174, 0.627) & $<0.001$ \\ % 5e\text{-}05
Optimal Power selection                & 0.464 & (0.228, 0.700) & $<0.001$ \\ %9.2e\text{-}06
Separate analysis & 0.417 & (-0.045, 0.878) & 0.084 \\
\bottomrule
\end{tabular*}
\begin{tablenotes}
\item Estimate: Risk difference between treatment and control from two-sample test for equality of proportions with continuity correction. 
95\% CI: confidence interval from two-sample test for equality of proportions with continuity correction.
$p$-value: Obtained from Fisher's Exact test with one-sided $\alpha$=0.025
\end{tablenotes}
\end{table*}
\end{center}

\clearpage

\section{Design-stage quantities (binary endpoint; Beta(1,1) for treatment; robust MAP for control)}
\label{sup:PoS}
When planning a hybrid RCT the observed rates of the historical control data are already available. Therefore, one can already condition on the observed results when calculating operating characteristics such as conditional type 1 error rates or power. The following section derives how these quantities can be calculated by using binomial likelihood. These formulas were used to calculate the conditional operating characteristics for the case study presented in Section 4.1 in the main manuscript.

Let $\pi_t^\star,\pi_c^\star\in(0,1)$ denote the \emph{assumed true} response rates for design evaluation, 
$n_t,n_c$ the planned sample sizes, and $\gamma$ the posterior success threshold (e.g.\ $\gamma=0.975$).
The treatment prior is $\pi_t\sim\mathrm{Beta}(1,1)$ and the control prior is a robust MAP derived from the selected historical trials $\Dhist$
\[
  \pi_c \sim w_R\,\mathrm{Beta}(1,1) + (1-w_R)\sum_{\ell=1}^K w_\ell\,\mathrm{Beta}(a_\ell,b_\ell),
  \qquad w_R\in(0,1),\; w_\ell>0,\; \sum_{\ell=1}^K w_\ell=1.
\]

\emph{Decision rule and boundary.}
Given counts $(y_t,y_c)$, the posterior probability of superiority decomposes as
\[
  \Pr(\pi_t>\pi_c \mid \Dnew,\Dhist) = \; 
  \widetilde w_R(y_c)\;\Pr\!\big[X>Y_0\big] 
  \;+\; \bigl(1-\widetilde w_R(y_c)\bigr)\sum_{\ell=1}^K \widetilde w_\ell(y_c)\;\Pr\!\big[X>Y_\ell\big],
\]
where $\Dnew=(y_t,y_c;\,n_t,n_c)$ as the notation of prospective trial data, and
\[
  X \sim \mathrm{Beta}(\underbrace{1+y_t}_{\alpha_1},\,\underbrace{1+n_t-y_t}_{\beta_1}),\quad
  Y_0 \sim \mathrm{Beta}(\underbrace{1+y_c}_{\gamma_0},\,\underbrace{1+n_c-y_c}_{\delta_0}),\quad
  Y_\ell \sim \mathrm{Beta}(\underbrace{a_\ell+y_c}_{\gamma_\ell},\,\underbrace{b_\ell+n_c-y_c}_{\delta_\ell}).
\]
Each term $\Pr[X>Y_\bullet]$ has the closed-form series \cite{pham-gia_bayesian_1993} when $\alpha_1\in\mathbb{N}$:
\[
  \Pr[X>Y] \;=\; \sum_{s=0}^{\alpha_1-1}
    \frac{B(\gamma+s,\;\beta_1+\delta)}{(\beta_1+s)B(\gamma,\delta)\,B(s+1,\beta_1)}.
\]
The updated mixture weights are based on component-wise marginal likelihoods
\[
  m_0(y_c) \propto \frac{B(1+y_c,\;1+n_c-y_c)}{B(1,1)},\qquad
  m_\ell(y_c) \propto \frac{B(a_\ell+y_c,\;b_\ell+n_c-y_c)}{B(a_\ell,b_\ell)},
\]
\[
  \widetilde w_R(y_c)=\frac{w_R\,m_0(y_c)}{w_R\,m_0(y_c)+(1-w_R)\sum_{j=1}^K w_j\,m_j(y_c)},\qquad
  \widetilde w_\ell(y_c)=\frac{w_\ell\,m_\ell(y_c)}{\sum_{j=1}^K w_j\,m_j(y_c)}.
\]
Define the data-dependent boundary as the largest treatment count not crossing the threshold,
\begin{equation}
d_1(y_c)=\max\{\,y_t\le n_t:\; \Pr(\pi_t>\pi_c \mid \Dnew,\Dhist)\le \gamma\,\}.
\label{eqS:boundary_rmap}
\end{equation}
so that success occurs if $y_t>d_1(y_c)$.

\emph{Conditional success probability (design stage).}
Given planning values $(\pi_t^\star,\pi_c^\star)$, the design-stage probability of meeting the success rule is
\begin{equation}
\label{eq:cp_binary_rmap}
\mathrm{CP}(\pi_t^\star,\pi_c^\star)
=\sum_{y_c=0}^{n_c}\binom{n_c}{y_c}(\pi_c^\star)^{y_c}(1-\pi_c^\star)^{n_c-y_c}
\sum_{y_t=d_1(y_c)+1}^{n_t}\binom{n_t}{y_t}(\pi_t^\star)^{y_t}(1-\pi_t^\star)^{n_t-y_t}.
\end{equation}

\emph{Design-stage operating characteristics (conditional on planning values).}
When evaluated at the null, $\pi_t^\star=\pi_c^\star=\pi$, \eqref{eq:cp_binary_rmap} yields the
\emph{conditional type I error rate}:
\[
\mathrm{T1E}_{\text{cond}}=\mathrm{CP}(\pi,\pi).
\]
When evaluated at an alternative with $\pi_t^\star>\pi_c^\star$, it yields the (design-stage)
\emph{conditional power}:
\[
\mathrm{Power}_{\text{cond}}=\mathrm{CP}(\pi_t^\star,\pi_c^\star).
\]

\emph{Probability of success (PoS, assurance).}
While conditional power ($CP$) assumes fixed true response rates
($\pi_t,\pi_c$), the probability of success averages $CP$ over the
design-stage priors for these parameters. For example, we use
$\pi_t \sim \mathrm{Beta}(1,1)$ and the robust MAP prior for $\pi_c$.
Formally,
\begin{equation}
\label{eq:pos_binary_rmap}
PoS
=\int_0^1\!\!\int_0^1 \mathrm{CP}(\pi_t,\pi_c)\;p_T(\pi_t)\;p_{\mathrm{rMAP}}(\pi_c)\;d\pi_t\,d\pi_c.
\end{equation}
Thus PoS represents the unconditional probability that the trial will declare
success, integrating over current uncertainty in treatment and control response
rates. In our implementation, \eqref{eq:cp_binary_rmap} and \eqref{eq:pos_binary_rmap}
correspond to \texttt{oc2S()} and \texttt{pos2S()} with
\texttt{decision2S(}$\gamma$\texttt{, 0, lower.tail=FALSE)}. The calculations use exact expressions, implemented in the \texttt{RBesT} package \cite{weber_applying_2021}.
\end{document}